\documentclass[10pt,aps,prb]{revtex4}

\usepackage{amssymb}
\usepackage{amsfonts}
\usepackage{amsmath}
\usepackage{graphicx}
\usepackage{dcolumn}
\usepackage{bm}

\begin{document}

\title{From {\ae}ther theory to Special Relativity
\footnote{Preliminary version of the article written for {\it
Handbook of Spacetime}, A. Ashtekar and V.Petkov (eds.),
Springer-Verlag GmbH (Heidelberg), in press.}}

\author{R{\scriptsize AFAEL} F{\scriptsize ERRARO}\footnote{ferraro@iafe.uba.ar.
Member of Carrera del Investigador Cient\'\i fico of CONICET.}\\}

\address{Instituto de  Astronom\'\i a y F\'\i sica
del Espacio, C.C. 67, Sucursal 28, 1428 Buenos Aires, Argentina\\
and
\\ Departamento de F\'\i sica, Facultad de Ciencias Exactas y
Naturales,\\ Universidad de Buenos Aires, Ciudad Universitaria,
Pabell\'on I, 1428 Buenos Aires, Argentina}

\begin{abstract}
At the end of the 19$^{\mathrm{th}}$ century light was regarded as
an electromagnetic wave propagating in a material medium called
\textit{ether}. The speed $c$ appearing in Maxwell's wave equations
was the speed of light with respect to the ether. Therefore,
according to the Galilean addition of velocities, the speed of light
in the laboratory would differ from $c$. The measure of such
difference would reveal the motion of the laboratory (the Earth)
relative to the ether (a sort of \textit{absolute motion}). However
the Earth's absolute motion was never evidenced.

Galileo addition of velocities is based on the assumption that
lengths and time intervals are \textit{invariant} (independent of
the state of motion). This way of thinking the spacetime emanates
from our daily experience and lies at the heart of Newton's
Classical Mechanics. Nevertheless, in 1905 Einstein defied Galileo
addition of velocities by postulating that light travels at the same
speed $c$ in any inertial frame. In doing so, Einstein extended the
\textit{principle of relativity }to the electromagnetic phenomena
described by Maxwell's laws. In Einstein's Special Relativity the
ether does not exist and the absolute motion is devoid of meaning.
The invariance of the speed of light forced the replacement of
Galileo transformations with Lorentz transformations. Thus,
relativistic length contractions and time dilations entered our
understanding of the spacetime. Newtonian mechanics had to be
reformulated, which led to the discovery of the mass-energy
equivalence.
\end{abstract}

\maketitle

\section{Space and time in Classical Mechanics}

Until 1915, when Einstein's General Relativity radically changed our way of
thinking, the \textit{spacetime} was regarded as the immutable scenery where the physical
phenomena take place. The laws of Mechanics, which describe the motion of a
particle subject to interactions, were written to work in this immutable
scenery. The form of these laws strongly depends on the properties
attributed to the spacetime. Classical Mechanics relies on the assumption
that distances and time intervals are invariant. This assumption, which
seems to be in agreement with our daily experience, leads to the Galilean
addition of velocities which prevents invariant velocities in Classical
Mechanics.

\subsection{Invariance of distances and time intervals}

Classical Mechanics --the science of mechanics founded by Newton--
considered that the space is properly described by Euclid's plane
geometry. Then there exist Cartesian coordinates ($x$, $y$, $z)$, so
the distance $d$ between two points placed at ($x_{1}$,$ y_{1}$,$
z_{1})$ and ($x_{2}$,$ y_{2}$,$ z_{2})$ can be computed by means of
the Pythagorean formula
\begin{equation}
\label{eq1}
d^{2}=(x_{2} -x_{1} )^{2}+(y_{2} -y_{1} )^{2}+(z_{2} -z_{1} )^{2}\quad .
\end{equation}
In addition, Classical Mechanics regards distances and time intervals as
\textit{invariant} quantities. Let us explain the meaning of this property with an example of
our daily life concerning the invariance of time intervals. Mario frequently
flies from Buenos Aires to Madrid; he knows that the journey lasts 12 hours
as measured by his watch. This time, Mario wants his friend Manuel to pick
him up at Madrid airport. When the fly is near to depart, Mario calls Manuel
who tells him that it is 9 am in Madrid. Then Mario asks Manuel to wait for
him at 9 pm in Madrid airport, just when the plane will be landing. This way
of arranging a meeting assumes that the time elapses in the same way both in
the plane and at earth. Of course, it seems to be a good assumption because
it effectively works in our daily life. We call \textit{invariant} a magnitude having the same
value in different frames in relative motion (as the plane and the earth in
the previous example). Classical Mechanics considers that not only time
intervals are invariant but the distances too. In particular, the length of
a body is assumed to be independent of its state of motion. We can
``verify'' this assumption in our daily life. For instance, we can measure a
train by spreading a tape measure on the train. The so obtained length will
seem to agree with a measure performed along the rail while the train is
traveling. Notice that measuring the length of a moving body requires some
care; the length is the distance between \textit{simultaneous} positions of the ends of the body.
In the case of the train, we can imagine that the rail is provided with
sensors detecting the stretch of rail the train takes up at each instant. We
can then determine the length of such stretch of rail by means of a tape
measure identical to the one used on the train.

The invariance of distances and time intervals are properties
supported by our daily experience. It could be said that space and
time look to us as separated concepts, and this separation seems not
to be affected by the choice of frame. This somehow naive way of
regarding the space and the time is a key piece in the construction
of Classical Mechanics. However, to what extent should we be
confident of our daily experience? Does our daily experience cover
the entire range of phenomena, or it is rather limited? Let us use a
familiar example to explain what we are trying to mean: we could
well believe that the earth surface is flat if just a little portion
of it were accessible to us. However, we realize that the earth
surface is nearly spherical by considering it at larger scales. In
this example, the scale should be comparable to the globe radius. In
the case of the behavior of distances and time intervals under
changes of frame, the scale in question is the relative velocity $V$
between the frames. How could we be sure that the invariance of
distances and time intervals is nothing but an appearance caused by
the narrow range of relative velocities $V$ covered by our daily
experience? As we will explain in Section IV, Einstein's Special
Relativity of 1905 abolished the invariance of distances and time
intervals on the basis of new physics developed in the second half
of the 19$^{\mathrm{th}}$ century.

\subsection{Addition of velocities}

Velocities are not invariant in Classical Mechanics. Let us consider
the motion of a passenger along a train traveling the rail at 100
m/s. The train and the earth are two possible frames to describe the
motion of the passenger; they are in relative motion at $V =$ 100
m/s. It is evident that the velocity of the passenger is different
in each frame. For instance, the passenger could be at rest on the
train, and thus moving at 100 m/s with respect to the earth. If the
passenger walks forward at a velocity of $u' =$ 1 m/s, then it
advances 1 meter on the train (as measured by a tape measure fixed
to the train) each 1 second (as measured by a clock fixed to the
train). Now, how fast does he/she move with respect to the earth?
The answer to this simple question depends on the properties of
distances and time intervals under change of frame. Since Classical
Mechanics assumes that distances and time intervals are invariant,
then we can state that the passenger advances 1 meter on the train
each 1 second as measured by a clock and a tape measure fixed
\textit{to the earth} (but otherwise identical to those fixed to the
train). Besides, in this frame also the train advances at the rate
of 100 meters each 1 second. Then, the passenger displaces 101
meters per second. Thus his/her velocity in the frame fixed to the
earth is $u =$ 101 m/s $= u' +V$. This \textit{addition of
velocities }is a direct consequence of the classical invariance of
distances and time intervals. It means that velocities are not
invariant in Classical Mechanics; they always change by the addition
of $V$. On the contrary, Einstein's Special Relativity will rebuild
our way of regarding the space and the time by postulating an
invariant velocity: the speed of light $c$ ($c =$ 299,792,458 m/s).
The postulate of invariance of the speed of light implies the
abandonment of our belief in the invariance of distances and time
intervals so strongly rooted in our daily experience. Therefore,
deep theoretical and experimental reasons should be alleged to
propose such a drastic change of mind. In fact, the idea of
invariance of the speed of light is theoretically linked to
Maxwell's electromagnetism and the principle of relativity, as will
be analyzed in Section III. Besides, at the end of the
19$^{\mathrm{th}}$ century there was enough experimental evidence
about the invariance of $c$. However those experimental results were
not correctly interpreted until Special Relativity came on stage.

The existence of an invariant speed provides us with a scale of reference to
understand why distances and time intervals seem to be invariant in our
daily life: according to Special Relativity, distances and time intervals
behave as if they were invariant when the compared frames (the train, the
plane, the earth, etc.) move with a relative velocity $V $\textless \textless
$c$. So, it is just an appearance; like the earth surface, that seems to be
flat if it is only explored in distances much smaller than the globe radius.

\subsection{Coordinate transformations}

An \textit{event} is a point in the spacetime. It represents a place in the space and an
instant of time; it is a ``here and now''. An event is characterized by 4
coordinates; we will use 3 Cartesian coordinates $x$, $y$, $z$, to localize the place
of the event plus its corresponding time coordinate $t$. Cartesian coordinates
are distances measured with rules along the Cartesian axes of the frame. The
coordinate $t$ is measured by clocks counting the time from an instant
conventionally chosen as the time origin.

Figure 1 shows two frames $S$ and $S'$ in relative motion; the $x$
and $x'$ axes have the direction of the relative velocity $V$. By
comparing distances in the frame $S$, we can state
\begin{equation}
\label{eq2} d_{OP\thinspace } \vert_{S}\ =\ d_{O{O}'\thinspace }
\vert_{S} +d_{{O}'P\thinspace } \vert_{S} \quad .
\end{equation}
In the frame $S$, the distance between $O$ --the coordinate origin
of $S$-- and the place $P$ is the $x$ coordinate of $P$:
$d_{OP\thinspace }|_{S} = x$. On the other hand, the distance
between the origins $O$ and $O'$ increases with time; if $V$ is
constant and the time $t$ in $S$ is chosen to be zero when both
origins coincide, then $d_{OO'\thinspace }|_{S}  = Vt$. Thus
\begin{equation}
\label{eq3} d_{{O}'P\thinspace } \vert_{S}\ =\ x\, -\, Vt\quad .
\end{equation}

\begin{figure}[htbp]
\centerline{\includegraphics[width=3.56in,height=1.42in]{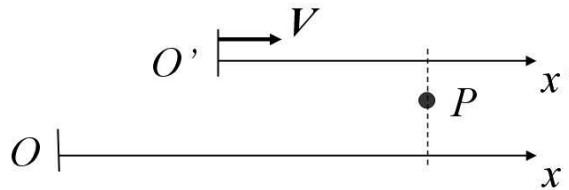}}
\caption{Frames $S$ and $S'$ moving at the relative velocity $V$.}
\label{fig1}
\end{figure}
We are not allowed to replace the left member with $x'$, since $x'
=d_{O'P\thinspace }|_{S'}$ . Classical Mechanics, however, assumes
that distances have the same value in all the frames. Thus, we
obtain the Galileo transformations:

\vskip.5cm \textbf{Galileo transformations}
\begin{subequations}
\begin{align}
\label{eq4a} {x}'&=x-Vt,\\
\label{eq4b} {y}'&=y,\\
\label{eq4c} {z}'&=z.
\end{align}
\end{subequations}
We have added the transformations of the Cartesian coordinates $y$, $z$
transversal to the relative motion of the frames. These are distances
between a given place and the straight line shared by the $x$ and $x'$ axes;
according to the classical invariance of distances, they are equal in $S$ and
$S'$.

The classical transformations of the coordinates of an event is
completed by considering the invariance of time intervals; so we
state that $t' = t$ (we are choosing a common time origin for $S$
and $S'$). Remarkably, the relation $t' = t$ also results from the
transformation (4.a) with the help of a simple physical argument: as
frames $S$ and $S'$ are on an equal footing, then the respective
inverse transformation should look like (4.a) except for the sign of
$V$ (if $S'$ moves towards increasing values of $x$ in $S$, then $S$
moves towards decreasing values of $x'$ in $S'$; thus the relative
velocity changes sign). Therefore
\begin{equation}
\label{eq7}
x={x}'+V{t}'.
\end{equation}
Then, by adding (4.a) and (5) one obtains
\begin{equation}
\label{eq1} {t}'=t.
\end{equation}

\vskip.5cm \textbf{Galileo addition of velocities}

A moving particle traces a succession of events in the spacetime.
This \textit{world-line} can be described by equations $x(t)$,
$y(t)$, $z(t)$, which are summarized in a sole vector equation for
the position vector \textbf{r}($t)$. According to Galileo
transformations (4), the position vector transforms as
\begin{equation}
\label{eq2} {\rm {\bf {r}'}}(t)={\rm {\bf r}}(t)-{\rm {\bf V}}t,
\end{equation}
where the invariance of time, $t' = t$, has also been used. By
differentiating Eq.~(7), it results the \textit{Galileo addition of
velocities}, i.e. the relation between the velocities of the
particle in two different frames due to the movement composition
with the relative translation between both frames:
\begin{equation}
\label{eq3} {\rm {\bf {u}'}}(t)={\rm {\bf u}}(t)-{\rm {\bf V}}.
\end{equation}
Velocities are not invariant under Galileo transformations. However,
the relative velocity between two particles is invariant:
\begin{equation}
\label{eq4} {\rm {\bf {u}'}}_{2} (t)-{\rm {\bf {u}'}}_{1}
(t)\;=\;{\rm {\bf u}}_{2} (t)-{\rm {\bf u}}_{1} (t).
\end{equation}

\vskip.5cm \textbf{Galilean invariance of the acceleration}

Since \textbf{V} is uniform, the differentiation of Eq.~(8) yields
the Galilean invariance of the acceleration:
\begin{equation}
\label{eq5} {\rm {\bf {a}'}}(t)={\rm {\bf a}}(t).
\end{equation}

\section{Relativity in Classical Mechanics}

Mechanics describes the motion of interacting particles by means of
equations governing the particle world-lines. These equations of
motion, together with the initial conditions, yield the coordinates
of particles as functions of time: $x(t)$, $y(t)$, $z(t)$. To write
the equations of motion we combine the laws of dynamics with the
laws of the interactions. Both types of laws must have the same form
in all the inertial frames. This is the principle of relativity in
Mechanics, which expresses that all the inertial frames are on an
equal footing. However, whether a given law consummates or not the
principle of relativity is a matter depending on the properties
attributed to the space and time.

\subsection{Newton's laws of dynamics}

Newton constructed the dynamics on the basis of three laws \cite{1}:

\vskip.5cm \textbf{First law (principle of inertia):} free particles
move with constant velocity (they describe straight world-lines in
spacetime).

\vskip.5cm \textbf{Second law: }a particle acted by a force acquires
an acceleration that is proportional to the force:
\begin{equation}
\label{eq6} {\rm {\bf F}}\ =\ m\,{\rm {\bf a}}.
\end{equation}
The proportionality constant $m$ is a property of the particle
called \textit{mass}. In terms of the \textit{momentum} ${\bf p}
\equiv m\, {\bf u}$, the law reads ${\bf F} = d{\bf p}/dt$.

\vskip.5cm \textbf{Third law (action-reaction principle): }two
particles interact by simultaneously exerting each other equal and
opposite forces.

First law is a particular case of the second law (the case ${\bf F
}= 0$); it establishes the tendency to perdurability as the main
feature of motion (as it was envisaged by Galileo \cite{2}, Gassendi
\cite{3} and Descartes \cite{4}, in opposition to the Aristotelian
thought). On the other hand, the second law becomes the particle
equation of motion, once the force is given as a function of
$\bf{r}$, $\bf{u}$, $t$, etc. Then, a law for the involved
interaction is also required (which can be gravitational,
electromagnetic, etc.). The third law implies the conservation of
the total momentum of an isolated system of interacting particles.
In fact, the reciprocal forces ${\bf F}_{12}$ and ${\bf F}_{21}$
between two particles $m_1$ and $m_2$ satisfies ${\bf F}_{12}  +
{\bf F}_{21} = 0$, since they are equal and opposite. If these are
the only forces on each particle, we can use the second law to
obtain $d({\bf p}_1 + {\bf p}_2)/ dt = 0$. Thus ${\bf p}_1+{\bf
p}_2$ is a conserved quantity. This argument can be extended to
prove the conservation of the total momentum of any isolated system
of particles.

Classical Mechanics allows for interacting forces at a distance.
They are derived from potential energies depending on the distances
between particles, which automatically provide interaction forces
accomplishing Newton's third law.

\subsection{Newton's absolute space}

Newton's fundamental laws of dynamics are not formulated to be used
in any frame. In fact, it is evident that the first law cannot be
valid in any frame, since a constant velocity \textbf{u} in a frame
$S$ does not imply a constant velocity ${\bf u}'$ in another frame
$S'$. This can be easily understood by considering cases where $S'$
rotates or accelerates with respect to $S$. However if $S'$
translates uniformly with respect to $S$, either the particle has
constant velocities \textbf{u}, ${\bf u}'$ in both frames or in none
of them. Galileo addition of velocities (8) is a particular example
of this general statement. In fact, Galileo transformations (4) were
obtained for two equally oriented moving frames; thus, they are in
relative translation (absence of relative rotation). Besides the
translation is uniform, since the velocity \textbf{V} is constant.
Thus ${\bf u}'$ is constant in (8) if and only if \textbf{u} is
constant.

Although the principle of inertia cannot be valid in any frame, at
least it is true that if it is valid in a frame $S$, then it will be
valid in any other frame $S'$ uniformly translating with respect to
$S$. Can we extent this statement to the second law? Second law
involves the particle acceleration. In Galileo transformations, the
acceleration is invariant. Besides, the forces in Classical
Mechanics depend on distances (like gravitational and elastic
forces) or relative velocities (like the viscous force on a particle
moving in a fluid, which depends on the velocity of the particle
relative to the fluid). Both the distances and the relative
velocities are invariant under Galileo transformations. In this way,
each side of second law (11) is invariant under changes of frames in
relative uniform translation. Therefore, the invariance of distances
and time intervals, which leads to Galileo transformations, is a key
piece in the Newtonian construction because it allows the second law
to be valid in a family of frames in relative uniform translation.
This is the family of \textit{inertial frames}, and this is the
content of the principle of relativity:

\vskip.5cm \textbf{Principle of relativity}

\textit{The fundamental laws of Physics have the same form in any
inertial frame.}

\vskip.5cm For instance, the same physical laws describe a free
falling body both in a plane and at the earth surface. The principle
of relativity in Classical Mechanics tells us that the state of
motion of the frame cannot be revealed by a mechanical experiment:
the result of the experiment will not depend on the motion of the
frame because it is ruled by the same laws in all the inertial
frames.

But how can we recognize whether a frame is inertial or not? We
could effectively recognize a particle in rectilinear uniform
motion; if we were sure that the particle is free of forces, then we
would conclude that the frame is inertial. However, Mechanics allows
not only for contact forces but for forces ``at a distance''. So how
can we be sure that a particle is free of forces? Newton was aware
of this annoying weakness of its formulation; he then considered
that the laws of Mechanics described the particle motion in the
\textit{absolute space}. Thus, the inertial frames are those fixed
or uniformly translating with respect to Newton's absolute space.

While the inertial frames are defined by their states of motion with
respect to Newton's absolute space, this (absolute) motion is not
detectable, since the principle of relativity puts on an equal
footing all the inertial frames; actually, only relative motions are
detectable. Absolute space in Classical Mechanics plays the
essential role of selecting the privileged family of inertial frames
where the fundamental laws of Physics are valid; but, surprisingly,
it is not detectable. In some sense absolute space \textit{acts},
because it determines the inertial trajectories of particles, but it
does not receive any reaction because it is immutable. Leibniz
\cite{5} criticized this feature of the Newtonian construction, by
demanding that Mechanics were aimed to describe relations among
particles instead of particle motions in the absolute space. In
practice, however, Newton's mechanics is successful because we can
choose frames where the non-inertial effects are weak or can be
understood in terms of \textit{inertial forces} that result from
referring the frame motion to another ``more inertial'' frame.

As advanced in Section I.B, Special Relativity will abandon the
invariance of distances and time intervals. Then, Galileo
transformations will be abandoned too. This means that Newton's
second law (11) and the character of fundamental forces will suffer
a relativistic reformulation. However the inertial frames will still
keep their privileged status devoid of a sound physical basis; this
issue will be only re-elaborated in General Relativity.

\section{The theory of light and the absolute motion}

In the second half of the 19$^{\mathrm{th}}$ century light was
regarded as electromagnetic mechanical waves governed by Maxwell's
laws. These waves were perturbations of a medium called ether; they
propagate at the speed $c$ relative to the ether. However, the ether
could not be evidenced, nor directly neither indirectly. Several
experiments did not succeed in revealing the Earth's motion relative
to the ether (a sort of absolute motion), and some forced hypothesis
about the interaction between matter and ether were introduced to
give account of these null results.

\subsection{The finiteness of the speed of light }

As mentioned in Section I.B, velocities are not invariant in
Classical Mechanics. Actually, only an infinite velocity would
remain invariant under Galileo addition of velocities (8). Are there
infinite speeds in nature? Many philosophers (Aristotle among them)
thought that the speed of light was infinite. The issue of whether
the speed of light was finite or infinite has been the object of
debate from the ancient times. In the 17$^{\mathrm{th}}$ century,
the question was still open. While Kepler and Descartes argued in
favor of an infinite speed of light, Galileo proposed a terrestrial
test that, however, was not suitable to determine such a large
speed. But at the end of 17$^{\mathrm{th}}$ century, contemporarily
to Newton's development of Mechanics, an answer came from the
Astronomy side.

In 1676 R{\o}mer \cite{6} noted that the time elapsed between the
observations of successive eclipses of Io --the closest of Jupiter's
great moons-- was larger when the Earth traveled its solar orbit
moving away from Jupiter and shorter when the Earth moved towards
Jupiter. R{\o}mer realized that such deviations in this otherwise
periodical phenomenon were the sign of a finite speed of light. In
fact, if the Earth were at rest, then we would observe one eclipse
each 42.5 hours (the orbital period of Io). However, if the Earth
moves away from Jupiter, the time between successive observations of
the emersions of Io from the shadow cone will be enlarged; this
happens because the light coming from the second emersion travels a
longer distance at a finite velocity to reach the Earth. This delay,
together with the length traveled by the Earth in 42.5 hours, led to
the first determination of the speed of light. By recording the
accumulative delay of many successive eclipses, R{\o}mer found that
the light traveled the diameter of the Earth's orbit in 22 minutes
(the actual value is 16 minutes) \cite{7}.

Fifty years later, Bradley \cite{8} discovered the aberration of
starlight. Bradley observed that the light coming from a star
suffers annual changes of direction in the frame translating with
the Earth. The nature of these changes highly disturbed Bradley
because they unexpectedly differed from the stellar parallax he was
looking for (a tiny effect only measured one hundred years after).
Eventually, Bradley concluded that the \textit{stellar aberration}
discovered by him was a consequence of the vector composition (8)
between the speed of light and the Earth's motion around the Sun at
30 km/s. By measuring the aberration angle, Bradley obtained the
speed of light within an error of 1{\%} \cite{9}. In 1849 Fizeau
\cite{10} carried out the first terrestrial measurement of the speed
of light. Like any finite velocity, the speed of light is not a
Galilean invariant.

\subsection{The wave equation }

At the middle of the 19$^{\mathrm{th}}$ century the dispute about
the corpuscular or undulatory character of light seemed to be
settled in favor of the wave theory of light. The corpuscular model
sustained by Newton and many other scientists could not explain the
totality of the luminous phenomena. In 1821 Fresnel \cite{11}
completed his wave theory of light, so giving a finished
mathematical form to the undulatory model proposed by Huygens in
1678 \cite{12}. This theory included the concepts of amplitude and
phase to describe interference and diffraction; besides, the light
was presented as a transversal wave to explain the phenomena
concerning polarization. In 1850 Foucault \cite{13} measured the
speed of light in water, and verified the value $c$/$n$ ($n$ is the
refractive index) as predicted by the wave theory in opposition to
the corpuscular model.

At that time, the light waves were considered \textit{matter} waves
like sound or the waves on the water surface of a lake. Physics and
Mechanics were synonymous; so, any phenomenon was regarded as a
mechanical phenomenon, and light did not escape the rule. Matter
waves propagate in a material medium; they are but medium
oscillations carrying energy. In the simplest cases, they are
governed by the \textit{wave equation}
\begin{equation}
\label{eq7} \frac{1}{c_{w}^{2} }\,\frac{\partial^{2}\psi }{\partial
t^{2}}-\nabla ^{2}\psi \ =\ 0\ ,
\end{equation}
where $\psi (t, {\bf r})$ represents the perturbation of the medium
(for instance the longitudinal oscillations of density and pressure
when sound propagates in a gas, or the transversal displacement of a
string in a musical instrument). Any function $\psi = \psi ( x \pm
c_{w\thinspace }t)$ is a solution of the wave equation (12); it
represents a perturbation that travels in the $x$-direction, without
changing its form, at the constant speed $\pm c_{w}$ . The general
solution is a combination of solutions traveling in all directions.

The wave equation (12) is not written to be used in any inertial
frame. It only describes the wave propagation in a frame fixed to
the medium. In fact, the wave equation changes form under Galileo
transformations. Let us take the $x$-sector of the Laplacian
$\nabla^{\mathrm{2}}$ and write:
\begin{equation}
\label{eq1} \frac{1}{c_{w}^{2} }\,\frac{\partial^{2}}{\partial
t^{2}}-\frac{\partial ^{2}}{\partial x^{2}}\;\,=\;\,\left(
{\frac{1}{c_{w} }\,\frac{\partial }{\partial t}-\frac{\partial
}{\partial x}} \right)\,\left( {\frac{1}{c_{w} }\,\frac{\partial
}{\partial t}+\frac{\partial }{\partial x}}
\right)\;\,=\,\;4\;\frac{\partial }{\partial \xi }\;\frac{\partial
}{\partial \eta }\quad ,
\end{equation}
where $\xi \equiv c_{w\thinspace }t - x$, $\eta \equiv
c_{w\thinspace }t + x$ (or $c_{w\thinspace }t  = (\eta + \xi )/2$,
$x =(\eta -\xi )/2$). This shows that the wave equation would keep
its form in different inertial frames moving along the $x$-axis if
$c_{w}\thinspace t \pm x $ were proportional to $c_{w}\thinspace t'
\pm x'$; but this is not true in Galileo transformations (4), (6).
The fact that the equation governing mechanical waves is fulfilled
just in the frame where the medium is at rest does not imply the
violation of the principle of relativity. The medium is a physical
reason for privileging an inertial frame; furthermore, the Eq.~(12)
will be accomplished whatever be the inertial frame where the medium
is at rest. Actually, the wave equation for mechanical waves can be
obtained from the fundamental laws of Mechanics --which certainly
accomplish the principle of relativity-- under some assumptions
valid in the frame fixed to the medium. In this derivation, the
propagation velocity $c_{w}$ results from the properties of the
propagating media.

\subsection{The {\ae}ther theory }

In Fresnel's theory, light was a mechanical wave that propagates in
a medium called the \textit{ether luminiferous}, and $\psi $ was the
``velocity of the ethereal molecules''. The speed of light $c$ was a
property of the ether. To be the seat of transversal waves, the
ether should be an elastic material; it was strange that no
longitudinal waves existed in this elastic medium. Besides, to
produce such enormous propagation velocity, the ether should be
extremely rigid. The ether should fill the universe, because light
propagates everywhere. It was logical to think the ether as at rest
in Newton's absolute space; the ether became a sort of
materialization of Newton's absolute space.

But such omnipresent substance should produce other mechanical
effects, apart from the luminous phenomena. How can planets move
through the ether without losing energy? Would the ether penetrate
through the moving bodies without disturbing them or it would be
dragged by them? If air is pumped out of a bottle, then the sound
will cease to propagate inside the bottle; however, the light will
still propagate, meaning that the ether was not evacuated together
with the air (why?). The ether looked like an elusive intangible
substance without any other effect than being the seat of the
luminous phenomena.

\subsection{Maxwell's electromagnetism }

In 1873 Maxwell \cite{14} published his Treatise on Electricity and
Magnetism, where electricity and magnetism appeared as two parts of
a sole entity: the electromagnetic field. Maxwell's laws for the
electromagnetic field contained as particular cases the well known
electrostatic interactions between charges and magnetostatic
interactions between steady currents. But the very Maxwell's
achievement was to discover that \textit{variable} electric and
magnetic fields --\textbf{E} and \textbf{B}-- create each other.
This mutual feedback between electricity and magnetism generates
\textit{electromagnetic waves}. In fact, Maxwell's equations in the
absence of charges lead to wave equations (12), with the Cartesian
components of \textbf{E} and \textbf{B} playing the role of $\psi $.
In the electromagnetic wave equations the propagation velocity is
$c_{\mathrm{\thinspace }}= (\mu_{\mathrm{o}}\varepsilon
_{\mathrm{o}})^{\mathrm{-1/2}}$. In SI units, $\mu_{\mathrm{o}}$ is
chosen to define the unit of electric current, and $\varepsilon_{o}$
is experimentally determined through electrostatic interactions;
their values are $\mu_{\mathrm{o\thinspace }}= 4\pi \times
10^{\mathrm{-7}}$ NA$^{\mathrm{-2}}$,
$\varepsilon_{\mathrm{o\thinspace }}= 8.854187817 \times
10^{\mathrm{-12}}$
N$^{\mathrm{-1}}$A$^{\mathrm{2}}$m$^{\mathrm{-2}}$s$^{\mathrm{2}}$.
To Maxwell's surprise, the value of $c$ coincided with the already
measured speed of light; so Maxwell concluded that light was an
electromagnetic wave.

Maxwell conceived the electromagnetic waves as a mechanical
phenomenon in a propagating medium. Therefore, he believed that his
equations were valid in a frame fixed to the medium. The recognition
of light as an electromagnetic wave then identified the
electromagnetic medium with the luminiferous ether. On another hand,
the action of the field on a charge $q$ --the Lorentz force ${\bf F}
= q ({\bf E}+{\bf u}\times {\bf B})$ -- depended on the velocity
\textbf{u} of the charge. This velocity was regarded as the velocity
of the charge with respect to the ether (the charge absolute
velocity).

Differing from Classical Mechanics, Maxwell's electromagnetism will
fit the Special Relativity without changes. Einstein will defy the
classical viewpoint by considering that Maxwell's equations should
be valid in any inertial frame. If so, the speed of light would be
invariant (i.e., it would have the same value in any inertial
frame). To sustain this idea, Galileo transformations should be
replaced with transformations leaving invariant the speed of light;
this implies the abandonment of the classical invariance of
distances and time intervals. In Special Relativity, Maxwell's
electromagnetism will become a paradigmatic theory.

\subsection{The search for the absolute motion}

Although the ether resisted a direct detection, at least it could be
indirectly tested. In the second half of the 19$^{\mathrm{th}}$
century, several experiments were aimed to test the Earth's motion
with respect to the ether (the Earth's absolute motion). While $c$
was considered the speed of light in the frame fixed to the ether,
the speed of light in the Earth's frame should result from composing
$c$ with the Earth's absolute motion $V$, according to the Galilean
addition of velocities (8). Therefore, some of these experiments
were based on the time the light takes to travel a round-trip along
a straight path (the light comes back after being reflected by a
mirror). To exemplify the idea, we will choose the path to be
parallel to the (unknown) Earth's absolute motion. According to
Galileo addition of velocities, the speed of light in the Earth's
frame is $c-V$ when light goes, and $c+V$ when light comes back. If
$l$ is the length the light covers in each journey, then the total
time of the round-trip is
\begin{equation}
\label{eq14}t\ =\ \frac{l}{c-V}\, +\, \frac{l}{c+V}\ =\ \frac{2\
l/c}{1-\frac{V^2}{c^2}}\ .
\end{equation}
As can be seen, the Earth's absolute motion $V$ enters the result as
a correction of the second order in $V/c.$ A correction of even
order was, in fact, expectable because the traveling time (14) does
not change if the Earth's motion is reversed. To be conclusive, the
experiments should be able to detect at least a value $V/c\sim$
10$^4$. This is because the Earth orbits the Sun at 30 km/s $\cong$
10$^{-4\thinspace }c$; then, even if the Earth were at rest in the
ether when the experiment is performed, it would move at 60 km/s six
months later. Therefore, any experimental array based on the
traveling time (14) should reach a sensitivity of
10$^{\mathrm{-8}}$. Such strong constraint could be circumvented by
experimental arrays sensitive to the change $V\to  -V$; if so, the
result could be of the first order in $V/c$. This the case of the
experiment performed by Hoek \cite{15} in 1868, where the symmetry
$V\leftrightarrow  -V$ is broken because one of the stretches of the
round-trip was not in air but in water; in this stretch, the speed
$c/n$ replaces $c$ in Eq.~(14). However, Hoek's interferometric
device was not effective for determining the Earth's absolute
motion.

There were also two experiments, sensitive to the first order in
$V/c$, that involved Snell's law. In 1871 Airy \cite{16} measured
Bradley's stellar aberration with a vertical telescope filled with
water. Bradley had measured the annual variation of the aberration
angle produced by the Earth's orbit around the Sun. This variation
did not reveal the Earth's absolute motion \textbf{V}  but just the
changes of \textbf{V}. Airy's experiment, instead, took into account
that the aberration implied that the telescope was not oriented
along the direction of the light ray in the ether's frame. If
Snell's law were valid in the ether frame, then an additional
refraction would take place when the light entered the water in the
telescope. This additional refraction would change the view angle to
the star by a quantity of the first order in $V/c$. Nevertheless,
Airy's experiment did not reveal the Earth's absolute motion. Much
earlier, in 1810, Arago \cite{17} covered a half of the objective of
a telescope with a prism, to obtain a second image of the stars. To
see the image through the prism, the telescope direction had to be
corrected in an angle equal to the deviation angle of the prism.
Arago believed that the light refraction in the prism could depend
on the velocity of light relative to the prism, which results from
the vector composition (8) of the speed of light with the absolute
motion of the prism (i.e., the Earth's absolute motion). This effect
could be revealed by observing stars in several directions to get
different vector compositions. However, Arago did not notice any
change of the deviation angle.

Fresnel \cite{18} searched reasons for Arago's null result. In the
context of the ether theory, he found that the null result could be
explained, at the first order in $V/c$, by advancing a curious
hypothesis: an (absolute) moving transparent substance partially
drags the ether contained in its interior. The partial dragging is
such that the phase velocity of light --the displacement per unit of
time of the wave fronts--, as measured in the frame fixed to the
universal ether (rather than the ether inside the substance) is not
$c$/$n$ but
\begin{equation}
\label{eq3} u=\frac{c}{n}+\left( {\,1-n^{-2}} \right)\;{\rm {\bf
V}}\cdot {\rm {\bf \hat{{n}}}}\ ,
\end{equation}
where ${\rm {\bf \hat{{n}}}}$ is the propagation direction,
\textbf{V} is the absolute motion of the transparent substance and
$n$ is its refractive index. In practice, Fresnel's dragging
coefficient $f = 1 - n^{-2}$ caused the fulfillment of Snell's law
in the frame fixed to the transparent substance (at the first order
in $V/c$). Fresnel's hypothesis explained why Arago did not succeed
in his endeavor: the deviation angle of the prism was always the one
predicted by Snell's law, irrespective of the absolute motion of the
prism. Besides, it also explained the null result in Airy's
experiment because no additional refraction will be produced if
Snell's law is valid in the frame fixed to the telescope (in this
frame the ray of light and the telescope are equally oriented).
Moreover, the partial dragging (15) cancels out the first order
effects in the time (14) when one of the stretches is not in air but
in another transparent substance; so, it also explained Hoek's null
result (Hoek's device was not sensitive enough to test second order
effects).

Fresnel's partial dragging of ether was measured by Fizeau \cite{19}
in 1851. Since Special Relativity will reject the existence of the
ether, Fizeau's measurement will require a relativistic
interpretation. On the other hand, the fulfillment of Snell's law in
the frame fixed to the transparent substance is completely
satisfactory in Special Relativity, because that is the only
physically privileged frame. For a detailed analysis of the
experiments pursuing the absolute motion in connection with
Fresnel's hypothesis, see References \onlinecite{20},
\onlinecite{21}.

\subsection{Michelson-Morley experiment}

In 1881 Michelson designed an interferometer aimed to detect the
Earth's absolute motion. In Michelson's interferometer the light
traveled round-trips completely in air. So, the challenge was to
achieve a sensitivity of 10$^{\mathrm{-8}}$. Figure 2 shows the
scheme of Michelson's interferometer. The beam of light emitted by
an extensive source is split into two parts by a half-silvered glass
plate. After travelling mutually perpendicular round-trips, both
parts join again to be collected by a telescope where interference
fringes are observed (Fizeau's fringes \cite{22}). The fringes are
caused by a slight misalignment of the mirrors; this implies that
the images of the mirrors at the telescope form a wedge. The wedge
causes that rays 1 and 2 arrive at the telescope with a phase-shift
that changes according to the thickness of the wedge at the place
where the rays bounced. So, the phase-shift will be different for
each one of the rays in the beam; therefore, bright and dark fringes
will be observed at the telescope. Notice that $l$ and
$L_{\mathrm{\thinspace }}$do not need to be equal, but $2(l  - L)$
should be smaller than the coherence length of light to preserve the
interference pattern.

For each ray in the beam, the phase-shift between parts 1 and 2
determines whether they produce a bright or a dark fringe. This
phase-shift results from the times $t_{\mathrm{1}}$,
$t_{\mathrm{2}}$ the rays 1 and 2 employ to cover their respective
round-trips; these times depend on the distances $l$, $L$ and the
velocities ${\bf u}'_1$, ${\bf u}'_2$ of the rays in the laboratory.
${\bf u}'_1$, ${\bf u}'_2$ are the result of the vector composition
(8) between the speed $c$ in the ether frame and the Earth's
absolute motion \textbf{V}; ${u}'_1$, ${u}'_2$ are clearly
different, since the vector composition depends on the direction of
each ray. Moreover, if the interferometer were gradually rotated
then the velocities ${\bf u}'_1$, ${\bf u}'_2$ would gradually
change. In this way, the rotation of the interferometer would affect
the fringes: the position of the bright fringes would gradually
displace. Instead, if the interferometer were at rest in the ether,
then the fringes would not displace because rays 1 and 2 would
travel at the speed $c$ irrespective of the orientation of the
interferometer. Thus, the displacement of the fringes would be the
indication of the Earth's absolute motion.

\begin{figure}[htbp]
\centerline{\includegraphics[width=3.86in,height=2.81in]{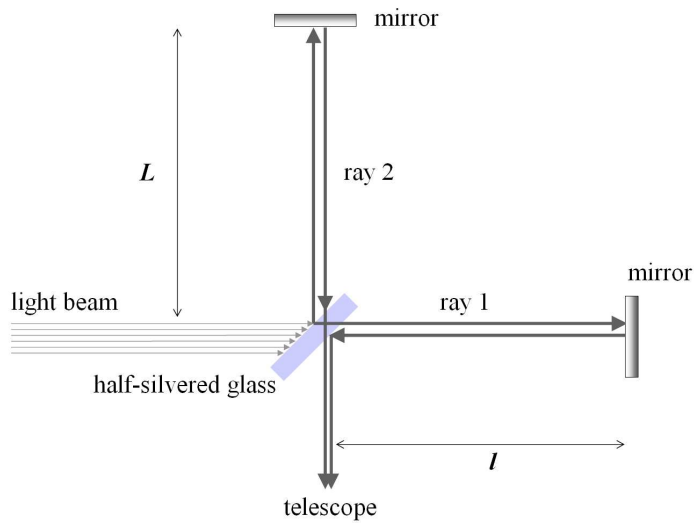}}
\caption{ Scheme of Michelson's interferometer.} \label{fig2}
\end{figure}

Let us compute the times $t_1$, $t_2$ when the arm $l$ is oriented
along the still unknown absolute motion \textbf{V}. In such case,
the ray 1 has speeds $c-V$, $c+V$, and the time $t_1$ is given by
Eq.~(14). On the other hand, the ray 2 is orthogonal to \textbf{V}
in the laboratory frame; so the vector composition to obtain the
value of ${\bf u}'_2$ is the one shown in Figure 3. As can be seen,
the ray 2 goes to the mirror and comes back with a speed
${u}'_2=\sqrt{c^2-V^2}$. Then, the round-trip along the arm $L$
takes a time
\begin{equation}
\label{eq1} t_{2} =\frac{2L/c}{\sqrt {\;1-\frac{V^{2}}{c^{2}}}
}\quad .
\end{equation}
The phase-shift is ruled by the time difference
\begin{equation}
\label{eq2} \Delta t_{0^{\circ }} =t_{1} -t_{2} =\frac{{2\,l}
\mathord{\left/ {\vphantom {{2\,l} c}} \right.
\kern-\nulldelimiterspace}
c}{\;1\;\;-\;\;\frac{V^{2}}{c^{2}}}\;-\;\frac{{2\,L} \mathord{\left/
{\vphantom {{2\,L} c}} \right. \kern-\nulldelimiterspace} c}{\sqrt
{\;1\;\;-\;\;\frac{V^{2}}{c^{2}}} }\ .
\end{equation}
\begin{figure}[htbp]
\centerline{\includegraphics[width=3.37in,height=2.25in]{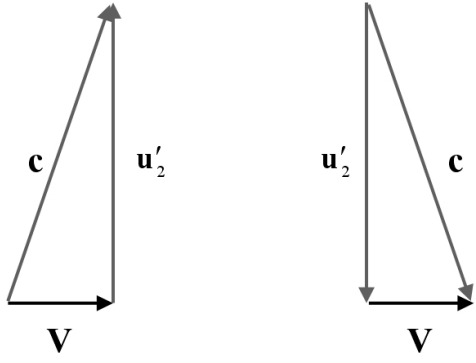}}
\caption{ Galilean composition of velocities for ray 2.}
\label{fig3}
\end{figure}

If the interferometer is rotated $90^\circ$, then the arm $L$
corresponding to the ray 2, will be aligned with \textbf{V}; so the
result will be
\begin{equation}
\label{eq3} \Delta t_{90^{\circ }} =t_{1} -t_{2} =\frac{{2\,l}
\mathord{\left/ {\vphantom {{2\,l} c}} \right.
\kern-\nulldelimiterspace} c}{\sqrt
{\;1\;\;-\;\;\frac{V^{2}}{c^{2}}} }-\frac{{2\,L} \mathord{\left/
{\vphantom {{2\,L} c}} \right. \kern-\nulldelimiterspace}
c}{\;1\;\;-\;\;\frac{V^{2}}{c^{2}}}\;\quad .
\end{equation}
Although the Earth's absolute motion \textbf{V} is unknown, a
gradual rotation will make the interferometer to pass through these
two extreme values separated by a right angle. Thus a displacement
of the fringes will be observed, in connection with the change of
$t_{\mathrm{1}}-t_{\mathrm{2}}$ given by
\begin{equation}
\label{eq4} \Delta t_{90^{\circ }} -\Delta t_{0^{\circ }}
=\frac{2\,}{\;c}\;(l+L)\;\left[ {\frac{1}{\sqrt
{\;1\;\;-\;\;\frac{V^{2}}{c^{2}}}
}-\frac{1}{\;\;1\;\;-\;\;\frac{V^{2}}{c^{2}}}}
\right]=-\frac{l+L}{\;c}\,\;\frac{V^{2}}{c^{2}}+O(V^{4}c^{-4}).
\end{equation}
This change is equivalent to the displacement of $N = c\ |\Delta
t_{90^{\circ}}-\Delta t_{0^{\circ} }|/ \lambda =$ ($l + L)/\lambda
\times V^2/c^2$ fringes ($\lambda$ is the light wavelength).

After a failed attempt in 1881, Michelson joined Morley to improve
the experimental sensitivity. In 1887 they possessed an
interferometer whose arms were 11 m long (this was achieved by means
of multiple reflections in a set of mirrors). Then, it was expected
at least a result of $N \cong$ 0.4. However, no displacement of
fringes was observed \cite{23,24,25}. Michelson was convinced that
the null result meant that the Earth carried a layer of ether stuck
to its surface. If so, the experiment would have been performed at
rest in the local ether, which would explain the null result. Lodge
\cite{26} tried to confirm this hypothesis by unsuccessfully looking
for effects due to the ether stuck to a fast rotating wheel. In a
revival of the corpuscular model, Ritz \cite{27} then proposed that
light propagates with speed $c$ relative to the source. This
hypothesis combined with other assumptions about the behavior of
light when reflected by a mirror (\textit{emission theories}) does
explain the null result of Michelson-Morley's experiment with a
source at rest in the laboratory, but is refuted by a varied body of
experimental evidence \cite{28,29,30}.

\subsection{FitzGerald-Lorentz length contraction}

Lorentz thought that Michelson-Morley's null result could be
understood in a very different way. He considered that a body moving
in the ether suffered a length contraction due to its interaction
with the ether. The interaction would contract the body along the
direction of its absolute motion \textbf{V}, but the transversal
dimensions would not undergo any change. In fact, if the contraction
factor $\sqrt {\;1-V^{2}c^{-2}} $ is applied to $l$ in Eq.~(17) and
$L$ in Eq.~(18) (i.e., the dimensions along the absolute motion
direction in each case), then both time differences will result to
be equal, and the expression (19) will vanish. This Lorentz's
proposal of 1892 \cite{31} had been independently advanced by
FitzGerald \cite{32} three years before. This proposal did not mean
the abandonment of the belief in the invariance of lengths. The
contraction was a dynamical effect; it depended on an objective
phenomena: the interaction between two material substances. The
contraction should be observed in any frame, and all the frames
should agree about the value of the contracted length.

The idea that light was a material wave (i.e., the idea that
Maxwell's laws were written to be used only in the ether frame) and
the belief in the invariance of distances and time intervals lead
Physics to a blind alley. While complicated dynamical explanations
were elaborated to interpret experimental results, like Fresnel's
partial dragging of ether and FitzGerald-Lorentz length contraction
caused by the ether, the experimental results were not so
complicated; they just said that the absolute motion cannot be
detected. However, unless Physics get rid of some classical
misconception, such a reasonable conclusion will not fit with its
theoretical body.

\section{Einstein's Special Relativity}

In 1905 Einstein postulated that ``\textit{the same laws of
electrodynamics and optics will be valid for all frames of reference
for which the equations of mechanics hold good}'' \cite{33}. In this
way, Einstein proclaimed that Mawell's electromagnetism does not
possess a privileged system; Maxwell's laws can be used in any
inertial frame. Thus, Einstein raised Maxwell's laws to the status
of fundamental laws satisfying the principle of relativity (as
stated in Section II.B). In doing so, Einstein closes the
possibility of detecting the state of motion of an inertial frame by
electromagnetic means. The ether does not exist; the electromagnetic
waves are not material waves. The inertial frames are not endowed
with a property \textbf{V} (its absolute motion or the ``ether
wind''); only the velocity describing the relative motion between
inertial frames makes physical sense. Besides, the Snell's law is
valid in the frame where the refracting substance is at rest,
whatever this frame is.

An immediate consequence of the use of Maxwell's laws in any
inertial frame is that light in vacuum propagates at the speed $c$
in any inertial frame; $c$ is an invariant velocity (``\textit{light
is always propagated in empty space with a definite velocity c which
is independent of the state of motion of the emitting body}''
\cite{33}). The existence of an invariant velocity implies that
Galilean addition of velocities is a classical misconception to be
got rid of; such step entails the revision of the classical belief
in the invariance of distances and time intervals.

\subsection{Relativistic length contractions and time dilations}

We will re-elaborate the transformations of spacetime coordinates
without prejudging about the behavior of distances and time
intervals, but subordinating them to the invariance of the speed of
light. Figure 4 shows a particle traveling between the ends of a
bar, as seen in the frame where the bar is fixed and the frame where
the particle is fixed. The relative motion bar-particle is
characterized by the sole velocity $V$. It is useful to call
\textit{proper length} $L_{o}$ the length of the bar at rest. Notice
that, since all inertial frames are on an equal footing, the length
of the bar will be $L_{o}$ in any inertial frame where the bar is at
rest. Instead, we could expect a different length $L(V)$ in a frame
where the bar moves lengthways at a relative velocity $V$. For this
reason, in Figure 4 the bar is represented with different lengths in
each frame. In the frame fixed to the bar (\textit{proper frame} of
the bar) the particle takes a time $\Delta t$ to cover the length
$L_{o}$; then, it is $V = L_{o\thinspace }$/$\Delta t$. On the other
hand, in the frame fixed to the particle, the ends of the bar take a
time $\Delta \tau $ to pass in front of the particle; then $V =
L_{\thinspace }$/$\Delta \tau $. We should not prejudge about the
nature of time; then, we are opening the possibility that the time
interval between the same pair of events be different in each frame.
It is also useful to call \textit{proper time} $\Delta \tau $ the
time between events as measured in the frame where the events occur
at the same place (if such a frame exists). In our case, the events
are the passing of each end of the bar in front of the particle;
they occur at the same place in the frame where the particle is
fixed. So, we have computed the same value of $V$ with lengths and
times measured in two frames that relatively moves at a velocity
$V$. Thus, we conclude that
\begin{equation}
\label{eq1} \frac{L_{o} }{L}=\frac{\Delta t}{\Delta \tau }\ .
\end{equation}
\begin{figure}[htbp]
\centerline{\includegraphics[width=3.40in,height=0.88in]{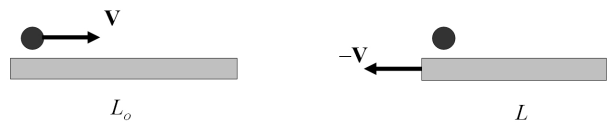}}
\caption{Relative motion bar-particle in the proper frames of the
bar (left) and the particle (right).} \label{fig4}
\end{figure}

Each side in Eq.~(20) could only depend on the relative velocity
between the considered frames. Then, the Eq.~(20) says that each
side is the same function of $V$:
\begin{equation}
\label{eq2} \frac{L_{o} }{L}=\gamma (V)\;,\quad \quad \quad \quad
\frac{\Delta t}{\Delta \tau }=\gamma (V).
\end{equation}
In Classical Physics $\gamma (V)$ is assumed to be 1. On the
contrary, in Special Relativity the value of $\gamma (V)$ will be
subordinated to the invariance of the speed of light. It should be
remarked that Eq.~(21) is not deprived of assumptions about the
nature of spacetime. In fact, the quotients $L_{o\thinspace }$/$L$
and $\Delta t_{\thinspace }$/$\Delta \tau $ could also depend on the
event of the spacetime where the measurements take place and the
orientation of the bar. Eq.~(21) actually assumes that the spacetime
is homogeneous and isotropic; these assumptions will be revised in
General Relativity.

On one hand, Eq.~(21) expresses the relation between the length $L$
of a bar moving at a velocity $V$ and its proper length $L_{o}$. On
the other hand, Eq.~(21) expresses the relation between the times
elapsed between two events as measured in the frame where they occur
at the same place (proper time $\Delta \tau )$ and other frame
moving at a velocity $V$ relative to the former one ($\Delta t)$. As
Eq.~(21) shows, both ratios are strongly interconnected.

The relations (21) are independent of the particular case examined
in Figure 4. To obtain $\gamma (V)$ we will now study a case
involving the speed of light, where the relations (21) will enter
into play too. Figure 5 shows a bar of proper length $L_{o}$
supporting at its ends a source of light and a mirror. Let us
consider the time elapsed between the emission of a pulse of light
from the source and its return to the source. Both events occur at
the same place in the proper frame of the bar; then, the proper time
$\Delta \tau $ is the time the light takes to cover the distance
2$L_{o}$ at the speed $c$:
\begin{equation}
\label{eq3} c\,\Delta \tau =2L_{o} \ .
\end{equation}
In another frame where the bar moves at a velocity $V$ (but light
still propagates at the speed $c)$, we will decompose the time
between events as $\Delta t  =  \Delta t_{going}  +  \Delta
t_{returning}$ . When light goes towards the mirror at the speed $c$
it covers the distance $L$ plus the displacement of the mirror $V
\Delta t_{going}$ . Instead, when light returns to the source it
covers the distance $L-V\Delta t_{returning}$ due to the
displacement of the source. Therefore,
\begin{equation}
\label{eq4} c\,\Delta \,t_{going} =L+V\,\Delta \,t_{going} \;,\quad
\quad \quad \quad c\,\Delta \,t_{returning} =L-V\,\Delta
\,t_{returning}\ .
\end{equation}

\begin{figure}[htbp]
\centerline{\includegraphics[width=3.80in,height=1.48in]{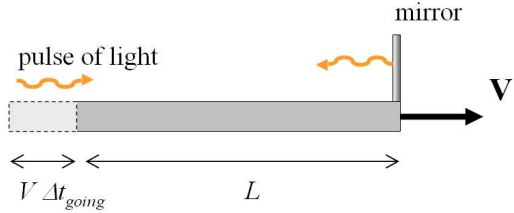}}
\caption{A light pulse traveling a round trip between the ends of a
bar, as regarded in the frame where the bar moves with velocity V.}
\label{fig5}
\end{figure}

Solving these equations for $c\Delta t_{going}$ , $c\Delta
t_{returning}$ one gets
\begin{equation}
\label{eq5} c\,\Delta {\kern 1pt}\,t\;\;=\;\;c\,\Delta \,t_{going}
+c\,\Delta \,t_{returning}
\;=\;\;\frac{c\;L}{c-V}\;\;+\;\;\frac{c\;L}{c+V}\;\;=\;\;\;\frac{2L}{1-\frac{V^{2}}{c^{2}}}\;\;.
\end{equation}
We divide Eqs.~(22) and (24), and use (21) for obtaining the
function $\gamma (V)$:
\begin{equation}
\label{eq6} \gamma (V)\;\;=\;\frac{1}{\sqrt
{\,1\;-\;\frac{V^{2}}{c^{2}}\;} }\ .
\end{equation}
Then, replacing in Eq.~(21) we get the expressions for the
relativistic \textit{length contraction} and \textit{time dilation}:
\begin{equation}
\label{eq7} L(V)\,\,=\;L_{o} \;\sqrt {\,1\;-\;\frac{V^{2}}{c^{2}}\;}
\;\;,\quad \quad \quad \quad \quad \Delta t_{V}
\,\,=\;\;\frac{\Delta \tau }{\sqrt {\,1\;-\;\frac{V^{2}}{c^{2}}\;}
}\ .
\end{equation}
Noticeably, the relativistic length contraction has the same form
proposed by FitzGerald and Lorentz to explain the null result of
Michelson-Morley experiment. However, its meaning is completely
different. Lorentz considered that the contraction was a dynamical
effect produced by the interaction between a body and the ether. For
Lorentz, $V$ in Eq.~(26) was the velocity of the body with respect
to the ether, and the contraction was measured in all the frames. In
Relativity, instead, the length contraction is a kinematical effect.
The bar looks contracted whatever be the frame where it moves at the
velocity $V$; besides, it has its proper length $L_{o}$ whatever be
the frame where the bar is at rest.

Length contractions and time dilations are not perceptible in our
daily life because we compare frames moving at relative velocities
$V $\textless \textless $c$. One of the first direct evidences of
this phenomenon came from measuring the length traveled by decaying
particles moving at a speed close to $c$, as compared to their
half-life measured at rest \cite{34}.

\subsection{Lengths transversal to the relative motion}

The device of Figure 5 is also useful to explore the behavior of the
dimensions transversal to the relative motion. Figure 6 shows the
device put in a direction orthogonal to the relative motion. Eq.
(22) is still valid in the proper frame of the bar. In a frame where
the bar transversally displaces at the velocity $V$, the ray of
light will travel along an oblique direction (this is nothing but
the aberration due to the composition of motions). When the pulse of
light goes towards the mirror, it covers in a time $\Delta
t_{going}$ the hypotenuse of a right triangle whose legs are
$V\Delta t_{going}$ and $L'$. Since the light travels at the speed
$c$ in any frame, we get
\begin{equation}
\label{eq1} (c\,\Delta \,t_{going} )^{2}={L}'\,^{2}\,+\,\,(V\,\Delta
\,t_{going} )^{2}\;.
\end{equation}
We remark the use of Pythagoras' theorem in this expression. This
means that we assume the space is endowed with a flat geometry; this
assumption will be revised in General Relativity. Due to the
symmetry of the path traveled by the light, it is $\Delta t =
2\Delta t_{going}$, then
\begin{equation}
\label{eq2} c\,\Delta t\;=\;\frac{2{L}'}{\sqrt
{\,1\;-\;\frac{V^{2}}{c^{2}}\;} }\;=\;\gamma (V)\;2{L}'\quad .
\end{equation}
We divide Eqs.~(22) and (28), and use (21) to get that transversal
lengths are invariant:
\begin{equation}
{L}'\;=\;L_{o}
\end{equation}

\begin{figure}[htbp]
\centerline{\includegraphics[width=3.77in,height=2.89in]{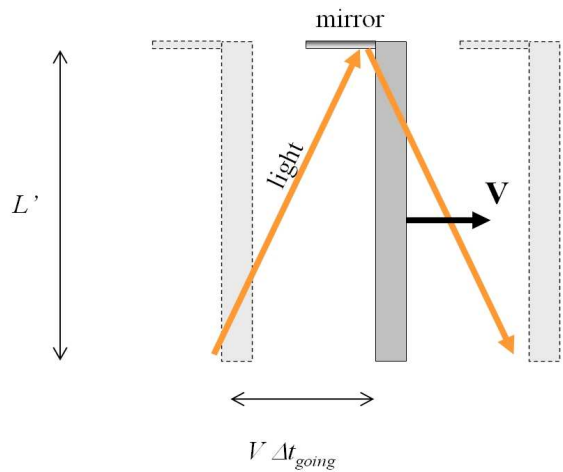}}
\caption{ The round trip of light between the ends of a bar, as
regarded in a frame where the bar displaces transversally at the
velocity $V$.} \label{fig6}
\end{figure}

\subsection{Lorentz transformations}

We are now in position of reanalyzing the transformation of the
Cartesian coordinates of an event. Let us come back to the Eq.~(3)
where the relation between $d_{O'P\thinspace }$\textbar $_{S}$ and
$x'$ is pending. By definition, the coordinate $x'$ is the distance
measured by a rule fixed to the frame $S'$: $x' = d_{O'P\thinspace
}$\textbar $_{S'}$ . This rule looks contracted in the frame $S$;
according to Eq.~(26) it is $d_{O'P\thinspace }$\textbar $_{S}  =
\sqrt {\;1-V^{2}c^{-2}}\ x'$. Therefore,
\begin{equation}
\label{eq3} x'\;\;=\;\;\gamma \;\left( {\,x\;-\;V{\kern 1pt}t}
\right)
\end{equation}
is the transformation that replaces (4.a). We can now reproduce the
argument of Section (I.C) to obtain the transformation of the time
coordinate of an event. Since frames $S$ and $S'$ are on an equal
footing, the inverse transformations have the same form, except for
the change $V\to $ $-V$. In particular, the inverse transformation
of (30) is
\begin{equation}
\label{eq4} x\;\;=\;\;\gamma \;\left( {\,{x}'\;+\;V{\kern 1pt}{t}'}
\right)\quad .
\end{equation}
Eq.~(30) can be replaced in (31) to solve $t'$ as a function of $t$,
$x$. Besides, due to the relativistic invariance of the transversal
lengths (see Eq.~(29)), the transformations (4.b), (4.c) remain
valid. Finally, we obtain the \textit{Lorentz transformations}:
\begin{subequations}
\begin{align}
\label{eq32a} ct'\;\;&=\;\;\gamma \;\left( {\,ct\;-\;\beta \,x}
\right)\ ,\\ \notag\\
\label{eq32b} x'\;\;&=\;\;\gamma \;\left( {\,x\;-\;\beta \,c{\kern
1pt}t} \right)\ ,\\ \notag\\
\label{eq32c} {y}'&=y\ ,\\ \notag\\
\label{eq32d} {z}'&=z\ ,
\end{align}
\end{subequations}
where $\beta \equiv  V/c$, $\gamma =\left( 1-\beta^2
\right)^{-1/2}$. Lorentz transformations (32) express the
relativistic transformation of the coordinates of an event, when the
inertial frame $S$ is changed for an equally oriented inertial frame
$S'$ that moves along the (shared) $x$-axis at the relative velocity
$V$. Notice that, since the transformation (32) is homogeneous, the
same event is the coordinate origin for $S$ and $S'$. Figure 7 shows
the lines $t' =$ \textit{constant} (i.e., $c t = \beta x +$
\textit{const}) and $x' =$ \textit{constant} (i.e., $c t =
 x/\beta +$ \textit{const}) in the plane $ct$ vs $x$.
Figure 7 also displays a ray of light passing the coordinate origin
and traveling in the $x-$direction; its world-line is a straight
line at $45^\circ$ because $\Delta x  =  c \Delta t$. Galileo
transformations (4) are the limit $c \to  \infty $ of Lorentz
transformations (32).

\begin{figure}[htbp]
\centerline{\includegraphics[width=3.43in,height=2.84in]{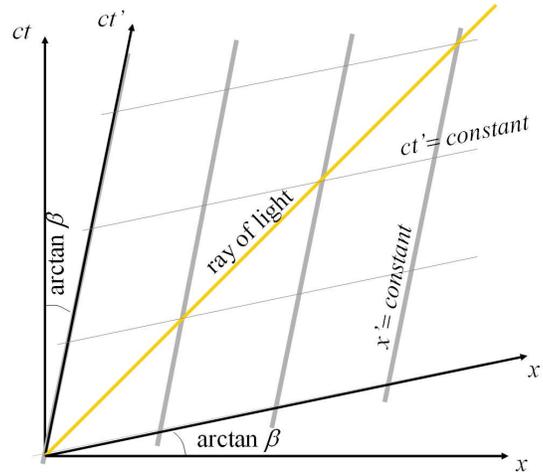}}
\caption{Coordinate lines of $S'$ in the plane $ct$ vs. $x$.}
\label{fig7}
\end{figure}

The transformations (32) were independently obtained by Lorentz
\cite{35,36} and Larmor \cite{37} as the linear coordinate changes
leaving invariant the form of Maxwell's wave equations (see also
Voigt \cite{38}). In fact, the \textit{null coordinates} $\xi \equiv
 c_{\thinspace }t - x$, $\eta \equiv c_{\thinspace }t + x$ transform as
$\xi ' = \gamma (1+\beta)\; \xi$, $\eta ' = \gamma (1-\beta )\;
\eta$, so leaving invariant the form of the wave equation (13) for
$c_{w} = c$. In other words, the d'Alembertian operator
\begin{equation}
\Box\ \equiv\ \frac{1}{c^2}\,\frac{\partial^2}{\partial t^2}\ -\
\nabla^2\, \label{eq33}
\end{equation}
is invariant under transformations (32). In 1905 Poincar\'{e} \cite{39}
underlined the group properties of relations (32) and called them
Lorentz transformations.

In 1905 Einstein re-derived the Lorentz transformations and gave to
$t'$ the rank of real time measured by clocks at rest in $S'$. In
Einstein's Special Relativity the physical equivalence of the
inertial frames, which is the content of the principle of
relativity, means that the fundamental laws of Physics keep their
form under Lorentz transformations rather than Galileo
transformations. Maxwell's laws accomplish this relativistic version
of the principle of relativity, once the transformations of the
fields are properly defined. Actually, Maxwell's electromagnetism is
the paradigm of a relativistic theory. The electromagnetic Lorentz
force is a typical relativistic force; its magnetic part depends on
the charge velocity relative to the inertial frame. But, which part
of the field is electric and which one is magnetic depends on the
frame as well; even if the force is entirely electric in a given
frame, it will have a magnetic part in another frame. On the
contrary, Classical Mechanics fulfilled the principle of relativity
under Galileo transformations; then, the Mechanics needed a
reformulation to accommodate to the relativistic meaning of the
principle of relativity.

\subsection{Relativistic composition of motions}

The composition of motions that replaces the Galilean addition of
velocities is obtained by differentiating the Eqs.~(32) and taking
quotients. Notice that
\begin{equation}
\label{eq5} dt\,'=\gamma \,(\,dt-\beta \,c^{-1}\,dx\,)=\gamma
\;\left( {1\;-\;\beta \,c^{-1}\;u_{x} } \right)\;dt\quad .
\end{equation}
Therefore,
\begin{subequations}
\begin{align}
\label{eq35a} u\,'_{x}\ &=\ \frac{dx\,'}{dt\,'}\ =\ \gamma \,\left(
{\frac{dx}{dt\,'}-V\,\frac{dt}{dt\,'}} \right)\ =\ \frac{u_{x}
\;-\;V}{1\;-\;\beta \,c^{-1}\;u_{x} }\;\;,\\ \notag\\ \label{eq35b}
u\,'_{y}\ =\ \frac{dy\,'}{dt\,'}\ &=\ \frac{\;\sqrt {\,1-\beta^{2}}
\;\;u_{y} \;}{1\;-\;\beta \,c^{-1}\;u_{x} }\;\;,\quad \quad
u\,'_{z}\ =\ \frac{dz\,'}{dt\,'}\ =\ \frac{\;\sqrt {\,1-\beta^{2}}
\;u_{z} \;}{1\;-\;\beta \,c^{-1}\;u_{x} }\;\;.
\end{align}
\end{subequations}
The procedure can be repeated to transform the accelerations.
Contrasting with Galilean transformations, the acceleration is far
to be invariant under Lorentz transformations.

Eqs.~(35) can be combined to get ${u'}^2 = {u'}_x^2 + {u'}_y^2 +
{u'}_z^2$; it is easy to verify that
\begin{equation}
\label{eq3}
1-\frac{{u}'^{\,2}}{c^{2}}\;\;=\;\;\frac{1-\beta^{2}}{(1-\beta
\,c^{-1}\,u_{x} )^{2}}\;\;\left( {1-\frac{u^{\,2}}{c^{2}}} \right).
\end{equation}
Since $\beta $ \textless 1 (otherwise Lorentz transformations should
be ill-defined), then both hand sides of Eq.~(36) have the same
sign. Therefore $u$ and $u'$ are both lower, equal or bigger than
$c$; this is an invariant property of the speed.

As an application of transformations (35), let us compute the speed
of light when light propagates in a transparent substance that moves
at the velocity $V$; then, $u'_{x}  =  c$/$n$, where $n$ is the
refractive index. We will use the inverse transformations to get
$u_{x}$ (i.e., we change $V$ for $-V$ in Eq.~(35.a)):
\begin{equation}
\label{eq4} \,u_{x}
\;\;=\;\;\frac{c}{n}\;\;\frac{1+\frac{nV}{c}}{1+\frac{\;V}{n\,c}}\;\;\approx
\;\;\frac{c}{n}\;\left( {1+\frac{nV}{c}} \right)\;\left(
{1-\frac{\;V}{n\,c}} \right)\;\;\approx \;\;\frac{c}{n}\;+\;\left(
{\;1-n^{-2}} \right)\;V.
\end{equation}
This result has the same form that Fresnel's partial dragging.
However, $V$ in Eq.~(37) is not the velocity of the transparent
substance with respect to the ether; it is the motion of the
transparent substance relative to an arbitrary inertial frame. What
Fizeau measured in 1851 was a relativistic composition of motions.

\subsection{Relativity of simultaneity. Causality}

Two events 1 and 2 (two points in the spacetime) are simultaneous if
they have the same time coordinate: $t_{\mathrm{1\thinspace }}=
t_{\mathrm{2\thinspace }}$. In Classical Physics the time is
invariant; so the simultaneity of events possesses an absolute
meaning. However, in Special Relativity $t_{\mathrm{1\thinspace }}=
t_{\mathrm{2}}$ does not imply $t'_{\mathrm{1\thinspace }}=
t'_{\mathrm{2\thinspace }}$. Then the simultaneity acquires a
relative meaning; it is frame-depending. In fact, the pairs of
events that are simultaneous in the frame $S$ lie on horizontal
lines ($t =$ \textit{constant}) in Figure 7; these lines cross the
$t' =$ \textit{constant} lines. Therefore the events simultaneous in
$S$ have different time coordinate $t'$ in $S'$.

To understand why the simultaneity is relative in Special
Relativity, let us consider a bar of proper length $L_{o}$ which is
equipped with a source of light at its center. In the proper frame
of the bar, a pulse of light will arrive simultaneously at both ends
of the bar, because it covers the same distance $L_{o}$/2 at the
same speed $c$ in both directions$.$ In another frame the bar is
moving but light still propagates at the speed $c$ in any direction.
Thus, the pulse will arrive before at the rear end of the bar
because this end moves towards the pulse of light. Then, the same
pair of events (the arrivals of the light to the ends of the bar) is
not simultaneous in a frame where the bar is moving. Moreover, since
which end is at rear depends on the direction of the motion (i.e.,
it depends on the frame), the temporal order of this kind of events
can be inverted by changing the frame.

\begin{figure}[htbp]
\centerline{\includegraphics[width=6.78in,height=2.69in]{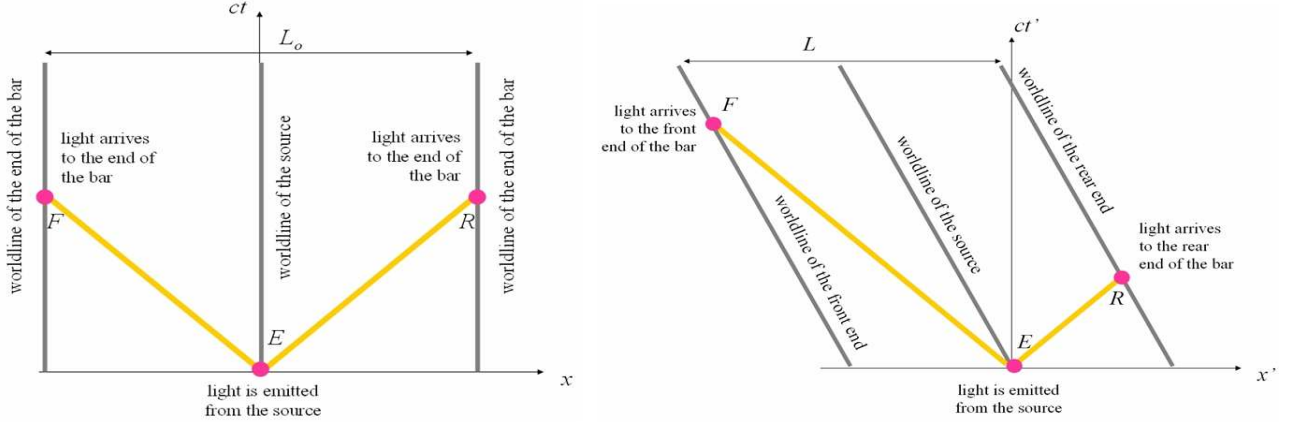}}
\caption{ a) In the proper frame of the bar, the pulses of light
arrive at the ends of the bar at the same time. b) In a frame where
the bar is moving, the light arrives at the rear end before than the
front end. In both frames the speed of light is $c$ (the rays of
light are lines at $45^\circ$).} \label{fig8}
\end{figure}

Figure 8 shows the world-lines of the ends of the bar and the pulses
of light both in the bar proper frame $S$ and a frame $S'$ where the
bar moves to the left (then $S'$ moves to the right relative to $S$,
so it is $\beta $\textgreater 0). In Figure 8a the ends of the bar
are described by vertical world-lines because the positions $x$ are
fixed. In Figure 8b the world-lines have the slope corresponding to
the velocity $-$\textbf{V} the bar has in the frame $S'$. In both
frames the light travels at the speed $c$. Events $R$ and $F$ are
simultaneous in the proper frame of the bar (Figure 8a), and they
occur at a distance $L_{o}$. Then, $\Delta t  = 0$, $\Delta x  =
-L_{o}$ ($\Delta t = t_{F} - t_{R}$, etc.). The time elapsed between
$R$ and $F$ in the frame $S'$ can be obtained by means of Lorentz
transformations. Since Lorentz transformations are linear, they are
equally valid for the differences of coordinates of a pair of
events. So, Eq.~(32.a) also means
\begin{equation}
\label{eq1} c\,\Delta {t}'\;\;=\;\;\gamma (c\,\Delta t\,-\beta
\,\Delta x)\ .
\end{equation}
Then it is $c\,\Delta {t}'\;=\;\gamma \,\beta \,L_{o}$ in Figure
8.b. This result could be also achieved by applying elementary
kinematics in the frame $S'$, and using the length contraction $L =
\gamma^{\mathrm{\thinspace \thinspace -1}} L_{\mathrm{o}}$ .

In any case, Eq.~(38) says that $\Delta t$ and $\Delta t'$ cannot be
both zero (apart from the case where the events are coincident).
Moreover, $\Delta t$ and $\Delta t'$ in Eq.~(38) could even have
opposite signs, which would amount to the inversion of the temporal
order of the events. This alteration of the temporal order in
Lorentz transformations would be acceptable only for pairs of events
without causal relation; otherwise it would constitute a violation
of causality. Remarkably, the violation of causality is prevented
because the speed of light cannot be exceeded in Special Relativity.
As it will be shown in Section V, $c$ is an unreachable limit
velocity for massive particles. Consistently, it is $V/c = \beta $
\textless 1 in Lorentz transformations. Therefore, those pairs of
events such that $|\Delta $\textbf{r}$|$\textgreater $c|\Delta t|$
cannot be in causal relation because neither particles nor rays of
light can connect them. For instance, in Figure 8 the events $R$ and
$F$ cannot be in causal relation because their spatial separation is
larger than their temporal separation. This property does not depend
on the chosen frame, as can be checked in the transformations (32)
or inferred from Eq.~(36). On the contrary, the pairs of events
having \textbar $\Delta $\textbf{r\textbar }$\le c \Delta t$ can be
causally connected. But in this case, it results that \textbar
$\beta \Delta x$\textbf{\textbar \textless }\textbar $\Delta
x$\textbf{\textbar }$\le  c \Delta t$. Thus \textbar $\beta \Delta
x$\textbf{\textbar } is not large enough to invert the temporal
order in Eq.~(38); so causality is preserved.

The relativity of simultaneity usually is the explanation to some
``paradoxes'' in Special Relativity. For instance, let us consider
two bars having the same length if compared at relative rest. Then,
if they are in relative motion, each one will appear shorter when
regarded from the proper frame of the other one. How could this make
sense? It makes sense because the length of a bar results from
comparing the simultaneous positions of its ends. Since the
simultaneity is not absolute in Special Relativity, then a length
measurement performed in $S$ is not consistent in $S'$.

\subsection{Proper time of the particle}

While those events having $|\Delta{\bf r}|$\textgreater $c\, |\Delta
t|$ admit a frame where they occur at the same time (or, moreover,
frames where their temporal order is inverted), those events having
$|\Delta{\bf r}|$\textless $c\, \Delta t$ admit a frame where they
occur at the same place. This is a consequence of the symmetric form
of Eqs.~(32.a) and (32.b). From a more physical standpoint, the
events having $|\Delta{\bf r}|$\textless $c\, \Delta t$ can be
joined by a uniformly moving particle. The proper frame of the
particle effectively realizes the inertial frame where both events
occur at the same place: the events occur at the (fixed) position of
the particle. These observations show that the concept of proper
time, as defined in Section IV.A, applies to pairs of events whose
spatial separation is smaller than the temporal separation.

In general, any moving particle causally connects events. Figure 9
shows the world-line of a particle that moves non-uniformly. Since
the world-line cannot exceed the angle of $45^\circ$ characterizing
the speed of light, any pair of events on the world-line of the
particle will satisfy $|\Delta{\bf r}|$\textless $c\, \Delta t$. Let
us consider two infinitesimally closed events, like those shown in
Figure 9 corresponding to the times $t$ and $t+$\textit{dt}. The
frame where these two events occur at the same place is the proper
frame of the particle moving at the speed $u(t)$. Let us rewrite the
Eq.~(36) with the help of Eq.~(34) to get
\begin{equation}
\label{eq2} \sqrt {\;1-\frac{{u}'^{\,2}}{c^{2}}}
\;\;d{t}'\;\;=\;\;\sqrt {\;1-\frac{u^{\,2}}{c^{2}}} \;\;dt\ .
\end{equation}
As is seen, this is a combination of speed and time of travel which
has the same value in any frame: it is invariant. By comparing with
Eq.~(26) one realizes that the invariant (39) is nothing but the
proper time elapsed between the infinitesimally closed events. In
other words, (39) is the time measured by a clock fixed to the
particle; it is the proper time of the particle:
\begin{equation}
\label{eq40} d\tau\ =\ \sqrt {\;1-\frac{u^{2}}{c^{2}}} \;dt\ =\
\gamma (u)^{-1}dt\ .
\end{equation}
\begin{figure}[htbp]
\centerline{\includegraphics[width=3.78in,height=2.69in]{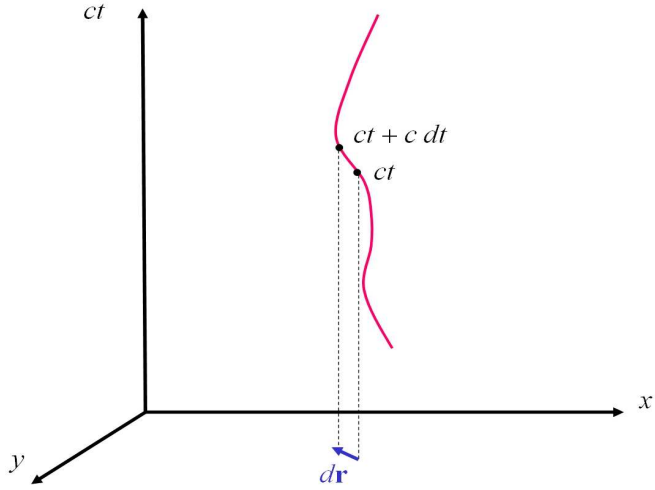}}
\caption{Two infinitesimally closed events belonging to the
world-line of nonuniformly moving particle. They are causally
connected: $|d{\bf r}| < c\, dt$.} \label{fig9}
\end{figure}

This expression can be integrated along the world-line to get the
total time measured by a clock that moves between a given pair of
causally-connectable events. Clearly, the integral depends on the
world-line the clock uses to join the initial and final events (it
depends on the function $u(t))$. It is easy to prove that the total
proper time is maximized along an inertial world-line. This result
is related to the so called \textit{twin paradox}. The paradox
refers to twin brothers who separate because one of them has a space
voyage. When they meet again, the ``inertial'' brother who remained
at the Earth is older than the astronaut. Actually this result is
not paradoxical; the brothers are not on an equal footing because
Special Relativity confers a privileged status to the inertial
frames.

\subsection{Transformations of rays of light}

Let us consider a monochromatic plane solution of the Eq.~(12) for
waves traveling at the speed of light:
\begin{equation}
\label{eq4} \psi \propto \exp \left[ {i\,\frac{2\pi \,\nu
}{c}\;(ct-{\rm {\bf \hat{{n}}}}\cdot {\rm {\bf r}})} \right]\quad ,
\end{equation}
where the unitary vector ${\rm {\bf \hat{{n}}}}$ is the propagation
direction, and $\nu $ is the frequency. Let us use the inverse
Lorentz transformations to rewrite the phase of the wave in terms of
coordinates in $S'$:
\begin{equation}
\label{eq5}
\begin{array}{l}
 \nu (ct-{\rm {\bf \hat{{n}}}}\cdot {\rm {\bf r}})\;=\;\nu \;\left( {\gamma
(ct'+\beta \,x')\;-\;n_{x} \gamma (x'+\beta \,ct')\;-\;n_{y}
y'\;-\;n_{z}
z'} \right) \\
 \\
 \quad \quad \quad \quad \;\,\,=\gamma \,(\,1-{\rm {\bf \hat{{n}}}}\cdot
{\rm {\bf V}}\,c^{-1})\;\nu \;ct'\;-\;\;\nu \left( {\gamma \,(n_{x}
-\beta
)\,x'\;+\;n_{y} y'\;+\;n_{z} z'} \right)\;. \\
 \end{array}
\end{equation}
Since the d'Alembertian operator (33) keeps the same form if
rewritten in coordinates of $S'$, the result (42) should be
reinterpreted as ${\nu }'(c{t}'-{\rm {\bf {\hat{{n}}}'}}\cdot {\rm
{\bf {r}'}})$. Therefore, one obtains:

\vskip.5cm\textbf{Doppler effect for light}

The frequency in the frame $S'$ is
\begin{equation}
\label{eq1} \nu '\;=\;\gamma \,(\,1-{\rm {\bf \hat{{n}}}}\cdot {\rm
{\bf V}}\,c^{-1})\;\nu\ .
\end{equation}
Factor $\gamma $ is absent in classical Doppler effect. It implies
that the frequency shift exists even if the propagation direction is
orthogonal to \textbf{V} (transversal Doppler effect) due to time
dilation. The first verification of the relativistic Doppler
frequency shift was made in 1938 \cite{40}.

\vskip.5cm\textbf{Light aberration}

Besides, it is ${n}'_{x} =(\nu /{\nu }')\,\gamma \,(n_{x} -\beta
)=(n_{x} -\beta )/(1-\beta \,n_{x} )$. If $\theta$ is the angle
between the ray of light and the $x$-axis, then it is $n_{x} =
\cos\theta $. Thus, the propagation direction transforms as
\begin{equation}
\label{eq2} \cos {\theta }'=\frac{\cos \theta -\beta }{1-\beta
\;\cos \theta }\quad .
\end{equation}
The aberration angle is $\alpha \equiv {\theta }'-\theta $; $\alpha
$ is very small whenever it is $\beta $\textless \textless 1. So, we
can approach $\cos {\theta }'=\cos (\theta +\alpha )\approx \cos
\theta -\alpha \,\sin \theta $. Besides, the right-hand side of
Eq.~(44) can be approached by $\cos \theta -\beta \,\sin^{2}\theta
.$ Therefore,
\begin{equation}
\label{eq1} \alpha \approx \beta \,\sin \theta ,
\end{equation}
which is the Galilean approach Bradley used to obtain the speed of
light from the annual variation of the starlight aberration.

\section{Relativistic Mechanics}

While the principle of inertia remains valid in Special Relativity,
instead Newton's second law has to be reformulated because it does
not satisfy the principle of relativity under Lorentz
transformations (forces and accelerations behave differently under
Lorentz transformations). The relativistic Mechanics can be
constructed from a Lorentz-invariant functional action that
reproduces the Newtonian behavior at low velocities. In Special
Relativity, energy and momentum are strongly related. The momentum
is conserved in any frame if and only if the energy is conserved
too. When particles collide, the conservation of the relativistic
energy takes the role of the classical mass conservation. However,
the relativistic energy is a combination of mass and kinetic energy;
so, mass can be converted in kinetic energy (or other energies, like
the electromagnetic energy associated with photons) and vice versa.
Classical interactions at a distance are excluded because the
relativity of simultaneity prevents non-local conservations of
energy-momentum. Instead, the interactions ``at a distance'' are
realized through mediating fields carrying energy-momentum that
locally interact with the particles.

\subsection{Momentum and energy of the particle}

The fulfillment of the principle of relativity under Lorentz
transformations can be achieved by starting from a Lorentz-invariant
functional action. In this way, it is guaranteed that different
inertial frames will agree about the stationarity of the action.
Thus, the same set of Lagrange dynamical equations will be valid in
all the inertial frames.

Let us start by building the action of a free particle. This action
not only has to be Lorentz invariant but must be equivalent to the
classical action when \textbar \textbf{u}\textbar \textless
\textless $c$. The (invariant) proper time along the particle
world-line (40) is the right choice for the functional action of the
free particle:
\begin{equation}
\label{eq2} S_{free} [{\rm {\bf r}}(t)]\;=\;-m\,c^{2}\;\int {d\tau }
\;=\;-m\,c^{2}\;\int {\,\sqrt {\,1-\frac{\vert {\rm {\bf u}}\vert
^{2}}{c^{2}}} } \;\,dt\;=\;-m\,c^{2}\;\int {\gamma (u)^{-1}\;dt} .
\end{equation}
When \textbar \textbf{u}\textbar \textless \textless $c$ the
Lagrangian $L = -mc^2(1-u^{\mathrm{2}}/c^2)^{\mathrm{1/2}}$ goes to
$L \approx  -mc^2 + (1/2)\, m\, u^2$. By differentiating the
Lagrangian $L$ with respect to \textbf{u} one gets the conjugate
momentum $m\gamma (u)$\textbf{u} of a free particle. One then
defines the \textit{momentum} of the particle as
\begin{equation}
\label{eq3} {\rm {\bf p}}\;\;\equiv \;\;m\;\gamma (u)\;{\rm {\bf
u}}\;\;=\;\;m\;\gamma (u)\;\frac{d{\rm {\bf
r}}}{dt}\;\;=\;\;m\;\frac{d{\rm {\bf r}}}{d\tau }
\end{equation}
(the last step results from Eq.~(40)), which goes to the classical
momentum $m$\textbf{u} when \textbar \textbf{u}\textbar \textless
\textless $c$.

Since $d\tau $ is invariant (see Eq.~(39)), the change of \textbf{p}
under Lorentz transformations emanates from the behavior of
$d$\textbf{r}. A Lorentz transformation mixes $d$\textbf{r} with
\textit{c dt}. Then \textbf{p} will be mixed with \textit{mc
dt}/$d\tau $, which is a quantity intimately related to the energy.
In fact, the Hamiltonian of the free particle is
\begin{equation}
\label{eq4} H\,\,=\,\,{\rm {\bf u}}\cdot {\rm {\bf
p}}-L\,\,=\,\,m\,\gamma (u)\,u^{2}+mc^{2}\gamma
(u)^{-1}\,\,=\,\,mc^{2}\gamma (u)\;\left(
{\frac{u^{2}}{c^{2}}+\gamma (u)^{-2}} \right)\,\,=\,\,mc^{2}\,\gamma
(u)\ .
\end{equation}
Then, we define the \textit{energy} of the particle as
\begin{equation}
\label{eq5} E\,\,\equiv \,\,m\,\gamma (u)\,c^{2}.
\end{equation}
The energy $E$ is a combination of \textit{energy at rest
mc}$^{\mathrm{2}}$ and kinetic energy. In fact, by Taylor expanding
(49) we obtain
\begin{equation}
\label{eq6}
E\,\,=\,\,mc^{2}\,+\,\,\frac{1}{2}m\,u^{2}\,\,+\,\,\mathellipsis
\,\,\equiv \,\,mc^{2}\,+\,\,T,
\end{equation}
where $T$ is the kinetic energy of the particle in Special
Relativity (at low velocities, it coincides with the classical
kinetic energy). Notice that the combination of (47) and (49) yields
\begin{equation}
\label{eq7} {\rm {\bf p}}\,\;=\,\;c^{-2}\,E\;{\rm {\bf u}}\ ,
\end{equation}
which says that the momentum is a flux of energy (as in
electromagnetism, where the density of momentum is proportional to
the Poynting vector).

Eq.~(40) can be used to replace $\gamma (u)$ in the energy (49); it
yields
\begin{equation}
\label{eq1} \frac{E}{c}\;\;=\;\;mc\;\frac{dt}{d\tau }\;.
\end{equation}
Then $E$ is proportional to the ratio of the time \textit{dt}
measured by frame clocks to the respective proper time of the
particle. As stated above, the invariance of $d\tau $ in Eqs.~(47)
and (52) implies that ($E$/$c$, \textbf{p}) transforms like ($c\,dt,
d{\bf r}$) under Lorentz transformations, i.e.:
\begin{subequations}
\begin{align} \label{eq53a} {E}'\;\;&=\;\;\gamma (V)\;(E-c\beta
\;p_{x})\;=\;\;\gamma (V)\;(E-{\rm {\bf V}}\cdot {\rm {\bf p}})\
,\\ \notag\\
\label{eq53b} {p}'_{x} \;&=\;\gamma (V)\;(p_{x} -\beta
\,c^{-1}E)\ ,\\ \notag\\
\label{eq53c} {p}'_{y} \;&=\;p_{y}\ ,\\ \notag\\
\label{eq53d} {p}'_{z} \;&=\;p_{z}\ .
\end{align}
\end{subequations}
$E^2$ and $c^2|{\bf p}|^2$ combine to yield the square particle
mass, an invariant result called the \textit{energy-momentum
invariant}:
\begin{equation}
\label{eq6} E^{2}-c^{2}\vert {\rm {\bf
p}}\vert^{2}\;=\;m^{2}c^{4}\gamma (u)^{2}-m^{2}c^{2}u^{2}\gamma
(u)^{2}\;=\;m^{2}c^{4}\,\left( {1-\frac{u^{2}}{c^{2}}}
\right)\,\gamma (u)^{2}\;=\;m^{2}c^{4}.
\end{equation}
Let us differentiate the Eq.~(54) to obtain
\begin{equation}
\label{eq7} E\;dE\;=\;c^{2}\;{\rm {\bf p}}\cdot d{\rm {\bf p}}\ ,
\end{equation}
or, replacing \textbf{p} with Eq.~(51):
\begin{equation}
\label{eq1} dE\;=\;{\rm {\bf u}}\cdot d{\rm {\bf p}}\;=\;d{\rm {\bf
r}}\cdot \frac{d{\rm {\bf p}}}{dt}\ ,
\end{equation}
which suggests that the force is associated with
$d$\textbf{p}/\textit{dt}. If so, the Eq.~(56) would express the
equality between the work of the force and the variation of the
energy. Notice that \textbf{F} $=  d$\textbf{p}/\textit{dt} implies
that the force is not parallel to the acceleration in general, due
to the term containing the derivative of $\gamma (u)$. Remarkably,
if the work goes to infinity, then the energy diverges and the
velocity $u$ in (49) goes to $c$. In this way, the speed of light is
an unreachable limit for the particle.

In electromagnetism, the interaction of a charge with a given
external field is described by adding the action (46) with the term
$S_{int} =-q\,\int {(\varphi -{\rm {\bf u}}\cdot {\rm {\bf A}})\,dt}
$, where $\varphi $ and \textbf{A} are the scalar and vector
potentials evaluated at the position of the charge. It can be proven
that the interaction action $S_{int}$ is Lorentz-invariant, as
required in Special Relativity. The variation of the action
$S_{free}  +  S_{int}$ leads to the equation of motion
\begin{equation}
\label{eq2} q\,({\rm {\bf E}}+{\rm {\bf u}}\times {\rm {\bf
B}})\;=\;\;\frac{d}{dt}\,\left( {m\,\gamma (u)\,{\rm {\bf u}}}
\right)\ ,
\end{equation}
where ${\bf E}=  -  \nabla \varphi  -
\partial{\bf A}/\partial t$ and ${\bf B }=\nabla
\times {\bf A}$. In Eq.~(57) we recognize the Lorentz force on the
left-hand side, and the derivative of the relativistic momentum (47)
on the right-hand side. In 1908 Bucherer \cite{41} observed the
movement of an electron in an electrostatic field, and obtained an
incontestable evidence of the validity of the relativistic dynamics
expressed in Eq.~(57). If the charge is initially at rest in a
uniform static field \textbf{E}, then we integrate the Eq. (57) to
get $(q/m){\bf E}\, t  = \gamma (u) {\bf u}$. So, $u$ goes to $c$
when $t$ goes to infinity.

\subsection{Photons}

In 1905 Einstein \cite{42} stated that the photoelectric effect
could be better understood by proposing that light interacts with
individual electrons by exchanging packets of energy $h\nu $ ($h$ is
Planck's constant and $\nu $ is the frequency of light). In this
way, the understanding of light-matter interactions required a new
concept where light shared characteristics of both wave and
corpuscle. In 1917 Einstein \cite{43} convinced himself that the
\textit{quantum of light} should be also endowed with directed
momentum, like any particle. The reality of the \textit{photon} was
confirmed by Compton's experiment in 1923 \cite{44}, where the
energy-momentum exchange between a photon and a free electron was
measured. The energy and momentum of photons traveling along the
${\rm {\bf \hat{{n}}}}$ direction,
\begin{equation}
\label{eq3} E_{photon} =\;\;h\nu \;,\quad \quad \quad \quad \quad
\quad {\rm {\bf p}}_{photon} =\;\;\frac{h\nu }{c}\;\;{\rm {\bf
\hat{{n}}}}\quad ,
\end{equation}
are those of a particle having zero mass (cf. Eq.~(54)) and the
speed of light (cf. Eq.~(51)). Lorentz transformations (53) for the
energy and the momentum (58) become the transformations (43) and
(44) for the frequency and the propagation direction of a ray of
light \cite{45}.

\subsection{Mass-energy equivalence}

In Relativity, the conservations of momentum and energy cannot be
dissociated. While the conservation of momentum comes from the
symmetry of the Lagrangian under spatial translations, the
conservation of energy results from the symmetry under time
translation. However space and time are frame-depending projections
of the spacetime. Space and time intermingle under Lorentz
transformations. Consequently, the conservation of momentum in all
the inertial frames requires the conservation of energy and vice
versa. This conclusion is evident in the transformations (53) where
energy and momentum mix under a change of frame; so, the momentum
would not be conserved in frame $S'$ if the energy were not
conserved in $S$. In sum, the conserved quantity associated to the
symmetry of the Lagrangian under spacetime translations is the total
energy-momentum.

In Classical Mechanics, instead, the transformation of the momentum
of the particle does not involve its energy. In fact, if Eq.~(8) is
multiplied by the mass, then the transformation ${\bf p}^\prime
 = {\bf p} - m{\bf V }$ is obtained. Thus, an isolated
system of interacting particles conserves the total momenta in all
the inertial frames irrespective of what happens with the classical
energy. Noticeably, $\Sigma \, {\bf p}^\prime$ is conserved whenever
$\Sigma\, {\bf p}$ is conserved because the total mass $\Sigma\, m $
is assumed to be a conserved quantity (classical principle of
conservation of mass). This is no longer true in Special Relativity.
For instance, let us consider the plastic collision between two
isolated particles of equal mass $m$. In the
\textit{center-of-momentum} frame the (conserved) total momentum
vanishes; so the particles have equal and opposite velocities $u$
before the collision. In the collision, the masses stick together
and remain at rest. If no energy is released, then the conservation
of energy implies
\begin{equation}
\label{eq4} 2\;m\;\gamma (u)\;c^{2}\;=\;M\,c^{2}\ ,
\end{equation}
where $M$ is the mass of the resulting body. Since $\gamma (u)$
\textgreater 1, then it is $M$ \textgreater 2$m $; in fact, the
resulting body contains the masses of the colliding particles and
their kinetic energies. In Einstein's words, ``the mass of a body is
a measure of its energy-content'' \cite{46}.

In general, the mass (energy at rest) of a composed system includes
not only the masses of its constituents but any other internal
energy as measured in the center-of-momentum frame. For instance, a
deuteron $D$ is constituted by a proton and a neutron. The deuteron
mass is lower than the addition of the masses of a free proton and a
free neutron; this evidences a negative binding energy among the
constituents. The \textit{mass defect} is $(m_{D}- m_{p}- m_{n}) c^2
 =  -$2.22 MeV. In general, when light nuclides merge into
a heavier nuclide (\textit{nuclear fusion}) some energy has to be
released to conserve the total energy. On the contrary, the mass of
a heavy nucleus is larger than the sum of the masses of its
constituents. Therefore, also there is a released energy in the
\textit{nuclear fission} of heavy nuclei. This dissimilar behavior
comes from the fact that the (negative) binding energy per nucleon
increases with the mass number for light nuclei but decreases for
heavy nuclei (the inversion of the slope happens at a mass number
around 60).

The kinetic energy can be used to create particles. For instance, a
neutral pion $\pi^0$ can be created in a high energy collision
between protons $p$; the reaction is $p +  p \to p +  p + \pi^0$.
This reaction can happen only if a \textit{threshold energy} is
reached to give account of the created particle. The neutral pion
has energy at rest (mass) of 134.98 MeV; then, in the
center-of-momentum frame the pion is created if each colliding
proton reaches the kinetic energy of 67.49 MeV. In such case, all
the kinetic energy is used to create the pion; the products remain
at rest, since no kinetic energy is left for the products, and the
total momentum is conserved. Therefore, the threshold energy of the
reaction in the center-of-momentum frame is equal to the energy at
rest of the products: $E_{threshold} = 2 m_{p} c^2 +  m_{\pi^0 }c^2
=$ 1876.54 MeV  $+$ 134.98 MeV. In this case, the energy balance is
(the particles are approximately free before and after the
reaction):
\begin{equation}
\label{eq5} 2\;m_{p} \;\gamma (u_{p} )\;c^{2}\;\,=\,\;2\;m_{p}
\,c^{2}+m_{\pi^{0}} \,c^{2}\quad \quad \Rightarrow \quad \quad
\gamma (u_{p} )\;\,=\,\;1+\frac{m_{\pi^{0}} }{2\;m_{p}
}\;\,=\,\;1+\frac{134.98}{1876.54}\;\,=\,\;1.072\quad ,
\end{equation}
which means that the velocity of the colliding protons in the
center-of-momentum frame is $u_{p} =$ 0.36 $c$. In another frame,
the threshold energy is higher because the products must keep some
kinetic energy to conserve the (non-null) total momentum. We can use
the Eqs.~(53) for transforming the total energy-momentum of the
system (since the transformations are linear, they can be used to
transform a sum of energies and momenta). In the center-of-momentum
frame the total momentum is zero; then Eq.~(53.a) says that
$E^\prime_{threshold} =  \gamma (V)\, E_{threshold}$. For instance
in the ``laboratory frame'' where one of the colliding protons is at
rest (i.e., $\gamma (V) = \gamma (u_{p}))$ it is
$E^\prime_{threshold} =$ 1.072 $E_{threshold}$; deducting the masses
of projectile and target, we obtain that the reaction is feasible if
the projectile reaches the kinetic energy of $T^\prime_{threshold} =
E^\prime _{threshold} - 2\, m_{p} c^2 =$ 279.67 MeV.

The previous example is a case of inelastic collision. A collision
is called \textit{elastic} if the particles keep their identities.
Thus, the masses (energies at rest) before and after the collision
are the same; so, the conservation of the energy of the colliding
free particles is equivalent to the conservation of the total
kinetic energy.

The interaction among charged particles can result in the release of
electromagnetic radiation. In such cases the radiation enters the
energy-momentum balance in the form of photons. For instance a pair
electron-positron annihilates to give two photons (the positron is
the anti-particle of the electron; they have equal mass but opposite
charge). In the center-of-momentum frame, the photons have equal
frequency and opposite directions to conserve the total momentum
(notice that at least two photons are needed to conserve the
momentum). If $u_{e}$ is the velocity of both particles in the
center-of-momentum frame, then the energy balance is
\begin{equation}
\label{eq1} 2\;m_{e} \;\gamma (u_{e} )\;c^{2}\;\,=\,\;2\;h\,\nu .
\end{equation}
Conversely, two photons can create a pair electron-positron. In this
case the threshold energy is equal to the mass of two electrons. So
the minimum frequency to create the pair in the center-of-momentum
frame is given by
\begin{equation}
\label{eq2} 2\;h\,\nu_{\min } \;\,=\,\;2\;m_{e} \;c^{2}\quad \quad
\Rightarrow \quad \quad \nu_{\min } \;\,=\;\,\frac{m_{e}
\;c^{2}}{h}\;\,=\;\,\frac{0.511\,\mbox{MeV}}{4.14\times
10^{-21}\,\mbox{MeV}\,\mbox{s}}\;\,=\;\,1.23\times
10^{20}\,\mbox{s}^{-1}\quad ,
\end{equation}
which is a frequency in the gamma-ray range of the electromagnetic
spectrum.

\vskip.5cm\textbf{Compton effect}

In 1923 Compton measured the scattering of X-rays by electrons in
graphite. X-ray photons have energies much larger than the electron
bound energies. So, the phenomenon can be studied as the elastic
collision between a photon and a free electron. In the frame where
the electron is initially at rest, its final momentum and energy are
\begin{equation}
\label{eq3} E_{e} =h\nu_{i} -h\nu_{f} +m_{e} c^{2}\;,\quad \quad
\quad \quad \quad {\rm {\bf p}}_{e} =h\nu_{i} \,c^{-1}\,{\rm {\bf
\hat{{n}}}}_{i} -h\nu_{f} \,c^{-1}\,{\rm {\bf \hat{{n}}}}_{f} \quad
,
\end{equation}
as results from compensating the changes of momentum and energy
suffered by photon and electron (in Eq.~(63) the labels $i $and $f$
allude to the initial and final states of the photon). The
replacement of these values in the electron energy-momentum
invariant (54) yields:
\begin{equation}
\label{eq4}
m_{e}^{2}c^{4}\;\,=\;\,E_{e}^{2}-p_{e}^{2}c^{2}\;\,=\;\,m_{e}
^{2}c^{4}+2\,h\;(\nu_{i} \;m_{e} c^{2}-h\nu_{i} \nu_{f} -\;m_{e}
c^{2}\;\nu_{f} +h\;\nu_{i} \,\nu_{f} \;{\rm {\bf \hat{{n}}}}_{i}
\cdot {\rm {\bf \hat{{n}}}}_{f} )\ .
\end{equation}
Eq.~(64) contains the relation between the ingoing and outgoing
photons. Let us call $\phi$ the angle between the initial and final
directions of propagation: $\;{\rm {\bf \hat{{n}}}}_{i} \cdot {\rm
{\bf \hat{{n}}}}_{f} =\cos \phi $. Then
\begin{equation}
\label{eq5} \frac{1}{h{\kern 1pt}\nu_{f} }\;-\;\frac{1}{h{\kern
1pt}\nu_{i} }\;\;=\;\,\frac{1}{m_{e} c^{2}}\;\,(\,1{\kern
1pt}-{\kern 1pt}\cos \phi \,)\quad \quad o\mbox{r}\quad \quad
\lambda_{f} -\lambda_{i} \;=\;\frac{h}{m_{e} c}\;\,(\,1{\kern
1pt}-{\kern 1pt}\cos \phi \,)\ .
\end{equation}
The quantity $\lambda_{C} \equiv  h/(m_{e\thinspace }c)  =$ 0.00243
nm is the electron \textit{Compton wavelength}. Eq.~(65) says that
the photon suffers a significant change only if its wavelength is
comparable to or smaller than the electron Compton wavelength (i.e.,
its energy is comparable to or larger than $m_{e\thinspace }c^2)$.

\subsection{Interactions ``at a distance''}

Interactions at a distance are allowed in Classical Mechanics; they
are described by potential energies depending on the distances
between particles, which automatically give equal and opposite
interaction forces accomplishing Newton's third law. Thus, although
the interaction forces change the momenta of the particles, these
changes cancel out by pairs at each instant; so the total momentum
of an isolated system of interacting particles is conserved.
Noticeably, the statement of Newton's third law cannot be translated
to Special Relativity, because the \textit{simultaneous}
cancellation at a distance has not an absolute meaning. In
particular, an interaction potential energy depending on the
distance between particles makes no sense in Relativity. Remarkably,
in electromagnetism the charges do not interact through such a
potential (apart from the static case). Instead, the interaction at
a distance is substituted for the \textit{local} interaction between
a charge and the surrounding electromagnetic field. This local
interaction entails the exchange of energy and momentum between
charge and field. The electromagnetic field carries momentum and
energy, which can be (partially) transferred to another charge at
another place. So, the isolated system conserving the total momentum
and energy is composed by the charges \textit{and} the
electromagnetic field. Conservation laws are local in Relativity.
The action governing an isolated system of charges and
electromagnetic field is the sum of the actions $S_{free\thinspace
}$ of the charges, the actions $S_{int\thinspace }$describing the
local interaction of each charge with the field at the place of the
charge (see Section V.A), and the invariant action of the
electromagnetic field $S_{field} =  \varepsilon_{o}/2 \int ({\bf
E}^2  -c^2\,{\bf B}^2)\,  d^3{\bf x}\ dt$.

\section{Conclusion}

As a theory about the structure of the spacetime, Special Relativity
is a framework to built theories in Physics: the laws governing any
physical phenomenon should be derived from Lorentz invariant
functional actions. In this way, the dynamical equations would
accomplish the principle of relativity under Lorentz
transformations. This requirement is enlighten in the
\textit{covariant} formulation to be developed in the next chapters.
Certainly, Maxwell's electromagnetism is a theory having the proper
behavior under Lorentz transformations. Also the field theories
describing subatomic interactions are built under relativistic
criterions. What about the theory of gravity? In Classical Physics,
gravity is a universal force proportional to the mass. The identity
between the \textit{gravitational mass} --the mass that measures the
strength of the gravitational interaction-- and the \textit{inertial
mass} --the mass in Eq.~(11)-- causes the motion of a
``freely-gravitating'' particle to be independent of its mass; it
just depends on the initial conditions. Einstein realized that this
fact opened the possibility of considering gravity not as a force
but as the geometry of the spacetime: the motion of a freely
gravitating particle would be the consequence of the geometry of the
spacetime. Special Relativity had revised the belief in the
invariance of lengths and times, but it still assumed that the space
was endowed with a frozen Euclid's flat geometry (which leads to the
Pythagoras' theorem we used in Eq.~(27)). Einstein went a big step
ahead to think that geometry could be a dynamical variable
determined by the distribution of matter and energy. Thus, Newton's
thought that matter is the origin of gravitational forces was
replaced by Einstein's idea that the energy-momentum distribution
determines the way of measuring the spacetime. In General
Relativity, geometry is governed by dynamical equations --the
Einstein equations-- fed by the energy and momentum located in the
spacetime; Special Relativity's geometry is just the geometry of an
empty spacetime. In General Relativity, the freely gravitating test
particles describe \textit{geodesics} of the spacetime geometry;
this is what a planet does when orbiting a star. Besides, when a
photon ascends a gravitational field, its frequency diminishes
because clocks go faster when the gravitational potential increases
(Eq.~(40) is no longer valid). The GPS system takes into account
this effect of gravity on the running of clocks to reach its highest
performance. So, the photon loses energy while ascending a
gravitational potential. This implies that its capacity of creating
mass decreases; but the so created mass is compensated for a larger
``potential energy''. In General Relativity the spacetime geometry
can evolve; thus we can interpret the cosmological data in the
context of an expanding universe. In sum, ten years after the birth
of Special Relativity, the concepts of space and time underwent a
new fundamental revision to tackle the relativistic formulation of
gravitational phenomena: Einstein's General Relativity has been
born.

\end{document}